\documentclass[numberedappendix]{emulateapj}

\usepackage{amssymb}

\shorttitle{Angular momentum - mass relation for haloes}
\shortauthors{Liao et al.}

\begin{document}
\title{Angular momentum - mass relation for dark matter haloes}
\author{Shihong Liao, Dalong Cheng, M. -C. Chu, and Jiayu Tang}
\affil{Department of Physics, the Chinese University of Hong Kong, Shatin, New Territories, Hong Kong SAR; shliao@phy.cuhk.edu.hk}

\begin{abstract}
We study the empirical relation between an astronomical object's angular momentum $J$ and mass $M$, $J=\beta M^\alpha$, the $J-M$ relation, using N-body simulations. In particular, we investigate the time evolution of the $J-M$ relation to study how the initial power spectrum and cosmological model affect this relation, and to test two popular models of its origin - mechanical equilibrium and tidal torque theory. We find that in the $\Lambda$CDM model, $\alpha$ starts with a value of $\sim 1.5$ at high redshift $z$, increases monotonically, and finally reaches $5/3$ near $z=0$, whereas $\beta$ evolves linearly with time in the beginning, reaches a maximum and decreases, and stabilizes finally. A three-regime scheme is proposed to understand this newly observed picture. We show that the tidal torque theory accounts for this time evolution behaviour in the linear regime, whereas $\alpha=5/3$ comes from the virial equilibrium of haloes. The $J-M$ relation in the linear regime contains the information of the power spectrum and cosmological model. The $J-M$ relations for haloes in different environments and with different merging histories are also investigated to study the effects of a halo's non-linear evolution. An updated and more complete understanding of the $J-M$ relation is thus obtained.
\end{abstract}


\section{INTRODUCTION}\label{s1}
The angular momentum - mass relation (or the $J-M$ relation) is a scaling relation between an astronomical object's angular momentum $J$ and its mass $M$. It was first noticed by \citet{brosche1963} that for a wide range of astronomical objects, from planet-satellite systems to super clusters, their $J$ and $M$ follow an empirical relation $J\propto M^{\sim 2}$. Later follow-up works confirmed this power law relation \citep[see][and references therein]{carrasco1982}. \citet{carrasco1982} then presented an updated version then of the $J-M$ relation  covering $\sim 30$ orders of magnitude in mass and $\sim 50$ orders of magnitude in angular momentum. \citet{fall1983} particularly studied the $J\propto M^{5/3}$ relation for spiral and elliptical galaxies. This relation was confirmed again by recent observational updates, e.g. \citet{romanowsky2012} and \citet{fall2013}. This $J\propto M^{5/3}$ relation was also observed for the cold dark matter haloes in cosmological N-body simulations; see e.g. \citet{efstathiou1979}, \citet{barnes1987}, \citet{sugerman2000}, etc.

This simple and universal relation from observation and simulation is unusual and demands an explanation. Here we briefly review two widely quoted explanations. The readers can refer to \citet{li1998} for another explanation from the global rotation of the universe.

\textit{Mechanical equilibrium}. This explanation usually appears in astronomical papers \citep[e.g.][]{ozernoy1967,carrasco1982}. When a galaxy (halo) becomes virialized, its rotational energy $K$ and gravitational energy $U$ are linked by the virial theorem, $2K+U=0$. Using $K\propto I\omega^2 \propto MR^2\omega^2$, $U \propto -GM^2/R$ and $M\propto R^3$, we can obtain $J\propto I\omega \propto MR^2\omega \propto M^{5/3}$. Here $I, \omega, R$ are the galaxy's (halo's) moment of inertia, average angular velocity and radius respectively, and $G$ is the gravitational constant. The key relation used in this explanation is the virial theorem, which implies that galaxies (haloes) are in mechanical equilibrium.

\textit{Arguments from the tidal torque theory}. In the tidal torque theory \citep[TTT,][]{stromberg1934,hoyle1949,peebles1969,doroshkevich1970,white1984}, a halo's angular momentum is induced by the tidal torques from the surrounding inhomogeneities, and thus has a dependence on the halo's moment of inertia and the tidal tensor. Since $I\propto MR^2$ and $M \propto R^3$, $J \propto I \propto M^{5/3}$ \citep[][]{peebles1969, white1994}. This scaling relation can also be addressed in detail by calculating the joint probability distribution of $J$ and $M$, $P(M,J)$; see e.g. \citet{catelan1996a}. From the ensemble results of TTT, \citet{catelan1996a} used the statistics of the initial density field to study $P(M,J)$, and found that $J$ is proportional to $M^{5/3}$ in the linear regime.

Although both explanations lead to the power index $5/3$ in the observed $J-M$ relation, they differ in the origin of the index. The mechanical equilibrium argument states that the $J-M$ relation is established in the virialized stage, while TTT claims it is found in the linear stage. Further investigations are needed to find the exact origin of the $J-M$ relation. Furthermore, the orbital-merger scenario \citep[][]{vitvitska2002,maller2002,peirani2004,donghia2007} shows that the nonlinear evolution has significant effects on the halo angular momentum after the turnaround stage. Whether the nonlinear evolution affects the $J-M$ relation is however not addressed. Also, the $J-M$ relation is tightly related to the evolution of angular momentum, which in turn depends on the initial perturbations and cosmological model (TTT). How do the power spectrum and cosmological model affect the $J-M$ relation? This is an interesting question that deserves to be answered.

In this paper we use N-body simulations to study the time evolution of the $J-M$ relation for protohaloes. Here a protohalo is defined as a clump of matter that is destined to end up as a halo at redshift $z=0$. Interestingly, $J\propto M^\alpha$ is found to be valid in the whole cosmological history, but with different $\alpha$ at different redshifts. This evolution behaviour enables us to test the two possible explanations mentioned above. The $J-M$ relation in the linear regime is shown to depend on the initial power spectrum and cosmological model. The dependences of the $J-M$ relation on the environment and merging history are also studied, to see the nonlinear evolution effects. We propose a three-regime scheme to explain the evolution of the $J-M$ relation, and give a more complete understanding for this relation. 

The structure of the paper is as follows. In Section \ref{s2} we briefly review the tidal torque theory, and derive the predictions for the $J-M$ in the linear regime. We describe our N-body simulation details, halo finders, environment classification method and merger tree constructions in Section \ref{s3}. Section \ref{s4} presents our results. The summary and discussion are given in Section \ref{s5}. The appendices summarise some numerical tests, including the simulation box size, resolution, halo finder, fitting method and smoothing schemes in TTT.

\section{\textit{J-M} RELATION IN THE LINEAR REGIME} \label{s2}
For the convenience of later discussion, we summarise some important steps of TTT from \citet{white1984} in Section \ref{s2s1}. We then derive the prediction for the $J-M$ relation in the linear regime in Section \ref{s2s2}.

\subsection{Tidal torque theory} \label{s2s1}
In the \textit{comoving} Eulerian coordinate ${\bf x}$, the total angular momentum of an object with respect to its center of mass ${\bf x}_\mathrm{cm}$ is
\begin{equation}\label{eqn01}
{\bf J}(t)=a^2\int_{V_\mathrm{cE}}\rho_\mathrm{com}({\bf x},t)({\bf x}-{\bf x}_\mathrm{cm})\times \dot{{\bf x}}d{\bf x},
\end{equation}
where $a$ is the scale factor, $V_\mathrm{cE}$ is the occupied region of the object in comoving Eulerian coordinate, and the comoving matter density can be expressed as
\begin{equation}\label{eqn02}
\rho_\mathrm{com}({\bf x},t) = \rho_0[1+\delta({\bf x},t)].
\end{equation}
$\delta({\bf x},t)$ is the dimensionless density contrast with respect to the comoving mean matter density $\rho_0$. With the Lagrangian perturbation theory, the mapping between comoving Eulerian coordinate ${\bf x}$ and \textit{Lagrangian} coordinate ${\bf q}$ is
\begin{equation}\label{eqn03}
{\bf x}={\bf q}+{\bf S}({\bf q},t).
\end{equation}
Here ${\bf S}({\bf q},t)$ is the displacement vector. The Jacobian transformation from ${\bf x}$ to ${\bf q}$ can be found by considering mass conservation. That is
\begin{equation}\label{eqn04}
|Q({\bf q})|=[1+\delta({\bf x},t)]^{-1}.
\end{equation}
With Equations (\ref{eqn02}), (\ref{eqn03}) and (\ref{eqn04}), the angular momentum in the corresponding Lagrangian region $V_\mathrm{L}$ can be expressed as
\begin{equation}\label{eqn05}
{\bf J}(t)=a^2\rho_0\int_{V_\mathrm{L}}({\bf q}-{\bf q}_\mathrm{cm}+{\bf S}-{\bf S}_\mathrm{cm})\times \dot{{\bf S}}d{\bf q}.
\end{equation}

In this paper, we only consider first order Lagrangian perturbations, i.e., the Zel'dovich approximation \citep{zeldovich1970}:
\begin{equation}\label{eqn06}
{\bf S}({\bf q},t)=-D(t)\nabla\psi({\bf q}),
\end{equation}
where $D(t)$ is the linear growth factor and $\psi({\bf q})$ is the gravitational potential. Higher order expressions can be found in \citet{catelan1996b}. 

Under the Zel'dovich approximation, the angular momentum is
\begin{equation}\label{eqn07}
{\bf J}(t)=-a^2\dot{D}(t)\rho_0\int_{V_\mathrm{L}}({\bf q}-{\bf q}_\mathrm{cm})\times \nabla\psi({\bf q})d{\bf q}.
\end{equation}

Further assuming the potential $\psi({\bf q})$ to be smooth in the region $V_\mathrm{L}$, we can approximate it using Taylor expansion at the centre of mass position up to second order
\begin{eqnarray}\label{eqn08}
\psi({\bf q})&\approx&\psi({\bf q}_\mathrm{cm})+\frac{\partial\psi({\bf q})}{\partial q_i}\bigg|_{{\bf q} = {\bf q}_\mathrm{cm}}(q_i-q_{\mathrm{cm},i}) \nonumber\\
&&+\frac{1}{2}\frac{\partial^2\psi({\bf q})}{\partial q_i\partial q_j}\bigg|_{{\bf q} = {\bf q}_\mathrm{cm}}(q_i-q_{\mathrm{cm},i})(q_j-q_{\mathrm{cm},j}),
\end{eqnarray}
where the Einstein summation convention is used. 

Substituting Equation (\ref{eqn08}) into Equation (\ref{eqn07}), we obtain the major result of the tidal torque theory:
\begin{equation}\label{eqn09}
J_i(t)=-a^2\dot{D}(t)\epsilon_{ijk}T_{jl}I_{lk},
\end{equation}
with the tidal tensor
\begin{equation}\label{eqn10}
T_{jl}=\frac{\partial^2\psi({\bf q})}{\partial q_j\partial q_l}\bigg|_{{\bf q} = {\bf q}_\mathrm{cm}},
\end{equation}
and the inertial tensor
\begin{equation}\label{eqn11}
I_{lk}=\rho_0\int_{V_\mathrm{L}}(q_l-q_{\mathrm{cm},l})(q_k-q_{\mathrm{cm},k})d{\bf q}.
\end{equation}

Equation (\ref{eqn09}) tells us that in the linear regime, the angular momentum of a protohalo depends on its shape (represented by $I_{lk}$) and the surrounding tidal torque (measured by $T_{jl}$) and evolves according to $a^2\dot{D}$, i.e. how the universe expands and how the perturbations grow. Specifically, the Levi-Civita symbol $\epsilon_{ijk}$ in Equation (\ref{eqn09}) implies that the angular momentum is produced due to the misalignment between the tidal tensor and inertial tensor. After the turnaround, the protohalo collapses to a virialized object (halo). TTT assumes that little angular momenta are gained or lost during this nonlinear process. TTT has been tested using N-body simulations with relatively good agreement. See \citet{sugerman2000} and \citet{porciani2002a,porciani2002b} for recent testings.

The temporal part of $J(t)$, $a^2\dot{D}$, depends on the cosmological model. Previous studies verified that, for a de Sitter universe, or for the matter dominated era in $\Lambda$CDM model, the halo angular momentum grows linearly with time as $a(t)^2\dot{D}(t) = t$ \citep[e.g.][]{white1984,sugerman2000}. We will test this temporal dependence for a quintessence dark energy model that has different expansion and structure growth rate as $\Lambda$CDM.

When calculating the tidal tensor $T_{jl}$, the potential field (or density field) is smoothed with a smoothing scale equal to the protohalo scale \citep{white1984},
\begin{equation}\label{eqn12}
T_{jl}=-\frac{1}{(2\pi)^3}\int k_j k_l \tilde{\psi}({\bf k}) \tilde{W}(kR_s) e^{i{\bf k}\cdot {\bf q}_\mathrm{cm}}d{\bf k}.
\end{equation}
Here, $\tilde{\psi}({\bf k})$ and $\tilde{W}(kR_s)$ are the Fourier transforms of the potential function and window function respectively. For the top-hat window function, the smoothing scale $R_s$ is usually set by $M=4\pi\rho_0 R_s^3/3$. As pointed out by \citet{white1984}, this smoothing process is needed to keep the validity of the Zel'dovich approximation used in TTT. The Zel'dovich approximation requires $|\delta^2| < 1$. However, inside a protohalo region, there may exist some smaller scale perturbations with $|\delta^2| > 1$ that need to be smoothed out. However, how to choose the value of $R_s$ is a nontrivial question. In Appendix \ref{as4}, we numerically test the choice of $R_s$ and show that the one usually adopted, $R_s=(3M/4\pi\rho_0)^{1/3}$, is the best choice.

The formalism of \citet{white1984} outlined above considers a random region $V_\mathrm{L}$ in the smooth density field which may not be a protogalaxy region. To study the angular momenta for density peaks, \citet{catelan1996a} calculated the ensemble average of angular momentum with respect to the potential field $\psi$,  
\begin{equation}\label{eqn13}
\langle{\bf J}^2\rangle_\psi = \frac{1}{15\pi^2}a^4 \dot{D}^2 (\mu_1^2 - 3\mu_2) \int dk k^6 P_\psi(k)\tilde{W}(kR_s)^2,
\end{equation}
where $\mu_1\equiv I_1+I_2+I_3, \mu_2\equiv I_1I_2 + I_1I_3 + I_2I_3$, and $I_1, I_2, I_3$ are eigenvalues of the inertial tensor $I_{ij}$. The term $\mu_1^2 - 3\mu_2$ depends on the statistical information of the density peaks. The potential power spectrum $P_\psi(k)$ is defined as $\langle\tilde{\psi}({\bf k}) \tilde{\psi}({\bf k'})\rangle_\psi = (2\pi)^3 \delta({\bf k}+{\bf k'})P_\psi(k)$.

\subsection{\textit{J-M} relation in the linear regime} \label{s2s2}
The $J-M$ relation is a statistical relation obtained from a large halo sample and has non-negligible scatterings. To
calculate the linear theoretical predictions, we use the ensemble results of TTT [Equation (\ref{eqn13})] and consider the simple scale-free models.

For a scale-free model with density power spectrum $P(k)=Ak^n$ in the linear regime, the potential power spectrum $P_\psi(k)$ is
\begin{equation}\label{eqn14}
P_\psi(k)= A(4\pi G\rho_0)^2k^{n-4}.
\end{equation}

Using a top-hat window function
\begin{equation}\label{eqn15}
\tilde{W}(kR_s)= 3\left[\sin(kR_s)-kR_s\cos(kR_s)\right]/(kR_s)^3,
\end{equation}
and $M=4\pi\rho_0 R_s^3/3$, we have
\begin{eqnarray}\label{eqn16}
\int dk k^6 P_\psi(k)\tilde{W}^2(kR_s) &=& 9A\left(4\pi G\rho_0\right)^2\left(\frac{4\pi}{3}\rho_0\right)^{1+\frac{n}{3}} \nonumber \\
& & \times M^{-1-\frac{n}{3}}I(n),
\end{eqnarray}
where $I(n)\equiv \int^\infty_0 dxx^{n-4}\left(\sin x -x\cos x\right)^2$.

Assuming $I_i = B_i M^{5/3}$, we obtain $\mu_1^2 - 3\mu_2 = B^2 M^{10/3}$, where $B_i$ and $B$ are constants that depend on protohaloes' shapes. Equation (\ref{eqn13}) becomes
\begin{eqnarray}\label{eqn17}
\langle |{\bf J}|\rangle_\psi &=& \left(\frac{48A}{5}\right)^{1/2} B G\rho_0 \left(\frac{4\pi}{3}\rho_0\right)^{\frac{1}{2}+\frac{n}{6}}I(n)^{1/2}a^2\dot{D} \nonumber \\
& & \times M^{\frac{7}{6}-\frac{n}{6}}.
\end{eqnarray}

Equation (\ref{eqn17}) implies that in the linear regime, for a model with scale-free $P(k)$, the $J-M$ relation has a constant power exponent 
\begin{equation}\label{eqn18}
\alpha =\frac{7}{6}-\frac{n}{6},
\end{equation}
and a time-dependent coefficient  
\begin{equation}\label{eqn19}
\beta(t)\propto a^2\dot{D}.
\end{equation}
Specifically, in the $\Lambda$CDM model, matter dominates in this regime, $D\sim a\sim t^{2/3}$, and thus $\beta(t)\propto t$. 

If we ignored the scale (or mass) dependence of $\int dk k^6 P_\psi(k)\tilde{W}(kR_s)^2$, then $\langle |{\bf J}|\rangle_\psi \propto (\mu_1^2 - 3\mu_2)^{1/2} \propto M^{5/3}$. This is how the previous arguments in TTT explain the observed $J-M$ relation. However, the smoothing scale $R_s$ in the smoothing potential is related to a protohalo's mass as $M=4\pi\rho_0 R_s^3/3$, and this introduces an additional mass dependence into TTT's predicted angular momentum. Therefore, when considering the $J-M$ relation in the linear regime, we cannot ignore this dependence. It leads to a deviation of $\alpha$ from $5/3$ in the linear regime. 

Equation (\ref{eqn18}) and Equation (\ref{eqn19}) are our predictions for the $J-M$ relation in the linear regime. We will test them in Section \ref{s4}.   

\section{NUMERICAL METHODS}\label{s3}

\subsection{N-body Simulation} \label{s3s1}
We used the public TreePM code GADGET2 \citep{springel2005} to perform all simulations. The initial conditions were generated using grid uniform particle distribution and Zel'dovich approximation. The simulations were divided into three groups: $\Lambda$CDM, scale-free and quintessence dark energy models.

The simulation parameters for $\Lambda$CDM model are summarized in Table \ref{t01}. It is known that a finite simulation box size could lower haloes' spins and affect the mass function \citep[see e.g.][]{bagla2005, power2006}. This might affect the $J-M$ relation and should be checked. We found that $L_\mathrm{box}\geq 100 h^{-1}$Mpc gave converged results (see Appendix \ref{as1}). In this paper, we only show the results of $\Lambda$CDM512b simulations which have a larger boxsize ($L_\mathrm{box}=200h^{-1}$Mpc) and thus better statistics of high mass haloes. Other simulations give similar results.

\begin{table*} 
\centering
  \caption{Simulation Parameters of the $\Lambda$CDM Model.}
  \begin{tabular} {@{}ccccccccccc@{}}
  \hline
     & & & & & & & & & Softening & \\ 
     Name & $\Omega_m$ & $\Omega_\Lambda$ & $\Omega_bh^2$ & $h$ & $N^3$ & $L_\mathrm{box}$ & $\sigma_8$ & $n_s$ & Length $\epsilon$ & Realizations\\
     & & & & & &  ($h^{-1}$Mpc) & & & ($h^{-1}$kpc) & \\ 
 \hline
   $\Lambda$CDM256 & 0.28 & 0.72 & 0.024 & 0.7 & $256^3$ & 100 & 0.8 & 0.96 & 3.0 & 10 \\
   $\Lambda$CDM512a & 0.28 & 0.72 & 0.024 & 0.7 & $512^3$ & 100 & 0.8 & 0.96 & 5.0 & 5 \\
   $\Lambda$CDM512b & 0.28 & 0.72 & 0.024 & 0.7 & $512^3$ & 200 & 0.8 & 0.96 & 20.0 & 10 \\
\hline
\end{tabular}
\label{t01}
\end{table*}

For scale-free simulations, we set up the initial conditions as in \citet{knollmann2008}. But instead of starting at the same scale factor $a$, our simulations began at different $a$ and stopped at the same $a=1$, in order to offer a direct comparison to our results from $\Lambda$CDM models. Especially, when compared to the $\Lambda$CDM or quintessence dark energy models with the time variable $t$, we use the corresponding $t$ in the Einstein-de Sitter cosmology for scale-free models. To normalize the power spectrum, we chose the characteristic nonlinear mass $M_\ast\approx 36000$ particles at $a=1$ for all simulations. The starting scale factor $a_i$ is set by requiring the integral power inside the box $\sigma^2_\mathrm{box}(a_i)=(2\pi)^{-3}\int d{\bf k}P(k,a_i)=0.15^2$ so that the simulation started with all scales in the linear regime. The normalization $A$ of the scale-free power spectrum $P(k)=Ak^n$ and $a_i$ for different simulations are listed in Table \ref{t02}.

\begin{table*}
 \centering
  \caption{Simulation Parameters of Scale-free Models.}
  \begin{tabular} {@{}ccccccc@{}}
  \hline
   Name & $n$ & $\Omega_m$ & $A$ & $a_i$ & $N^3$ & Realizations\\
 \hline
   SF-0.50 & -0.50 & 1.0 &3358.69 & 6.76E-4 & $256^3$ & 4\\
   SF-1.00 & -1.00 & 1.0 &1603.62 & 1.69E-3 & $256^3$ & 4\\
   SF-1.50 & -1.50 & 1.0 &674.74 & 4.36E-3 & $256^3$ & 4\\
   SF-2.00 & -2.00 & 1.0 &239.27 & 1.16E-2 & $256^3$ & 4\\
\hline
\end{tabular}
\label{t02}
\end{table*}

For the homogeneous dynamical dark energy simulation, we use the AS quintessence model \citep{albrecht2000}, which has a significant portion of dark energy in early times and thus a notably different growth factor $D(t)$ from the $\Lambda$CDM model (see Section \ref{s4s1}). We adopt the parametrization formula for quintessence dark energy's equation of state in \citet{corasaniti2003}, i.e.
\begin{equation}\label{eqn20}
w_\mathrm{Q}(a)=w^0_\mathrm{Q}+(w^m_\mathrm{Q}-w^0_\mathrm{Q})\times \frac{1+e^{\frac{a^m_c}{\Delta_m}}}{1+e^{-\frac{a-a^m_c}{\Delta_m}}}\times \frac{1-e^{-\frac{a-1}{\Delta_m}}}{1-e^{\frac{1}{\Delta_m}}},
\end{equation}
with parameters $w^0_\mathrm{Q}=-0.96, w^m_\mathrm{Q}=-0.01, a^m_c=0.53$ and $\Delta_m=0.13$. Other cosmological parameters in this simulation are the same as $\Lambda$CDM512b. We started an AS model simulation with the same initial conditions as a $\Lambda$CDM512b run. Therefore, the output differences give a direct and clean comparison between the growth of angular momenta in two cosmologies.

\subsection{Halo Identification}\label{s3s2}
We adopted the AMIGA Halo Finder \citep[AHF,][]{knollmann2009} to extract haloes in our simulation outputs. The virial overdensity parameter $\Delta_\mathrm{vir}(z)$ is set according to
\begin{equation}\label{eqn21}
\Delta_\mathrm{vir}(z)=18\pi^2+82x-39x^2,
\end{equation}
where $x=\Omega_m(z)-1$ \citep{bryan1998}. We have tested that the $J-M$ relation results are not sensitive to $\Delta_\mathrm{vir}$ for a wide range of its values. We excluded subhaloes in our analysis since subhaloes usually are tidally disrupted and their angular momenta vary violently. We have used the Friends-of-friends \citep[FOF,][]{davis1985} halo finder to cross check the AHF results, and their $J-M$ relation results were consistent with each other.

In order to determine the minimum particle number $N_\mathrm{min}$ to define a halo for angular momentum studies, we performed a resolution test and found that $N_\mathrm{min}= 200\sim 400$ is needed to obtain converged results (see Appendix \ref{as3}). In this paper, we choose conservatively $N_\mathrm{min}= 400$.

\begin{figure*}
\includegraphics[width=500pt]{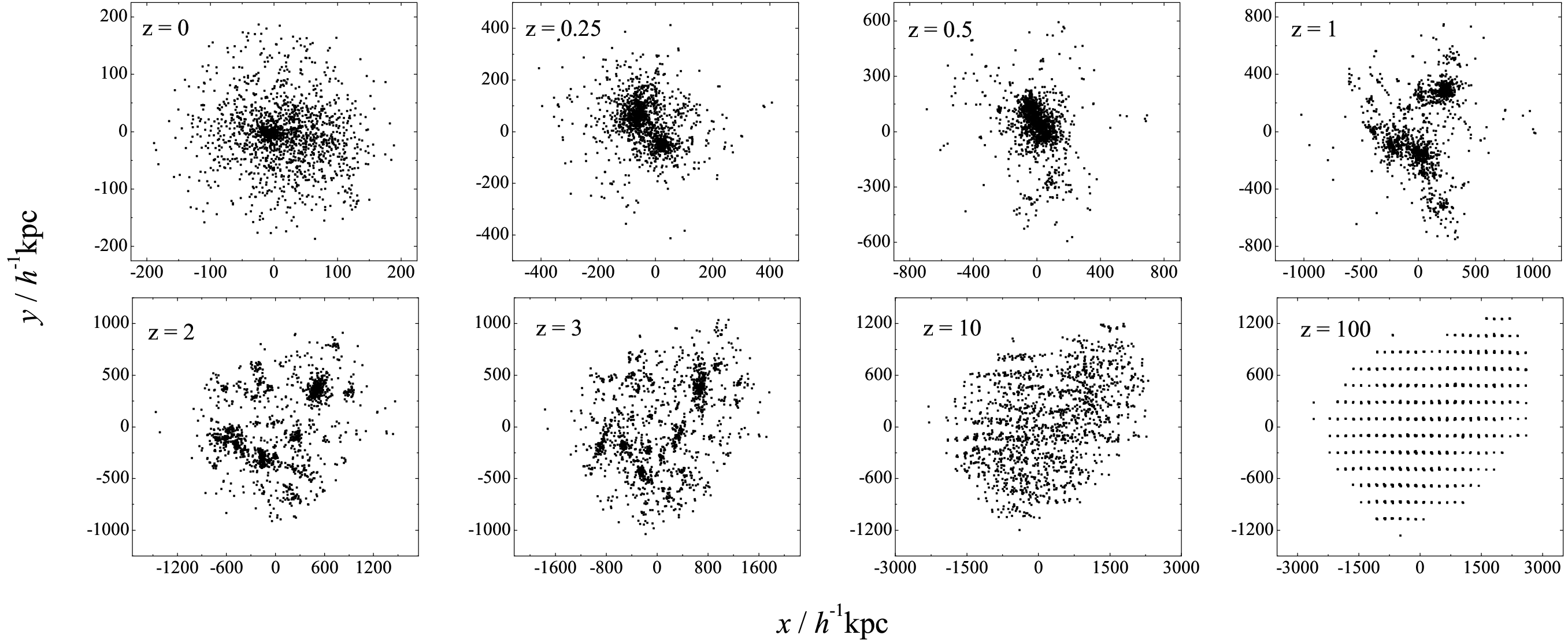}
\caption{Time evolution of a protohalo in the trace-back picture (projected on the $xy$ plane using comoving coordinate). The protohalo's position has been translated to keep its centre of mass at the origin. The figures show us how a clump of matter, with initial small perturbations, experiences inhomogeneous collapse and finally becomes a virialized halo.}
\label{f01}
\end{figure*}

To study the time evolution of the $J-M$ relation, we identified haloes at $z=0$ and traced the particles within these haloes back to the earlier time. Protohaloes are defined as the configurations of these particles in earlier time (see Figure \ref{f01}). 

A halo's angular momentum and mass are calculated as
\begin{equation}\label{eqn22}
{\bf J}(t)=\sum_i m_i[{\bf r}_i(t)-{\bf r}_\mathrm{cm}(t)]\times[{\bf v}_i(t)-{\bf v}_\mathrm{cm}(t)],
\end{equation} 
and
\begin{equation}\label{eqn23}
M=\sum_i m_i,
\end{equation}
respectively. Here the summation is over all particles within a protohalo. ${\bf r}_\mathrm{cm}(t)$ and ${\bf v}_\mathrm{cm}(t)$ are the centre of mass position and velocity. Notice that in this trace-back picture, the halo mass $M$ is a constant. 

We used two independent methods to fit the $J-M$ relation: All Points Fitting (APF) and Mass Bins Fitting (MBF). The details of these methods are described in Appendix \ref{as2}. They showed consistent results. In the text, if not mentioned, we only show results using the MBF method.

\subsection{Environment Classification}\label{s3s3}

We used the Hessian matrix method \citep{hahn2007} to classify the cosmic web. The Hessian matrix
\begin{equation}\label{eqn24}
H_{ij}({\bf r})\equiv\frac{\partial^2\rho_s({\bf r})}{\partial r_i \partial r_j}
\end{equation}
was calculated from the density field smoothed with a Gaussian kernel (smoothing scale $R_s = 2.1h^{-1}$Mpc). The eigenvalues of $H_{ij}({\bf r})$ are then calculated for each halo in its centre of mass position. A halo is classified as cluster/filament/sheet/void type if it has 0/1/2/3 positive eigenvalues (i.e. the classification threshold $\lambda_\mathrm{th}=0$).

Here, we use the density field to classify the cosmic web \citep[see also][]{zhang2009}. One can use other fields such as the potential field, velocity divergence field and velocity shear field \citep[e.g.][]{hahn2007,hoffman2007,cautun2013}. 

\subsection{Merger Trees}\label{s3s4}
In our simulations, there were 30 snapshots ranging from $z=5$ to $z=0$ with time intervals of $0.1 \sim 0.5$ Gyrs. To construct merging histories, we identify haloes in each of 30 snapshots with a minimum particle number of 20. Then, progenitor haloes in snapshot $n$ which merge to form a halo in the subsequent snapshot $n+1$ (target halo) are identified by locating particles of the target halo in haloes of snapshot $n$. We call the progenitor halo that contributes most particles to the target halo as ``mother'' and the ones contributing less as ``satellites''. Notice that there is no satellite for some haloes. It implies that these haloes increase their masses by small accretions. Also, for some haloes - especially high redshift and low mass ones - we may not be able to find their mothers, because their progenitors are too small to show up in our halo catalogue. We only use those haloes whose progenitors can be traced back to $z>2$.  

To study the dependence of the $J-M$ relation on the halo merging history, we divided all haloes at $z=0$ (with $N_\mathrm{min}=400$) into two groups, major merger (MM) and minor merger (mM), according to two parameters: the satellite-to-mother mass ratio $r_m$ (defined as the mass ratio between the largest satellite halo and the mother halo) and merger redshift $z_m$. If a merger event with $r_m \geq r_\mathrm{th}$ occurs for $z_m\leq z_\mathrm{th}$ ($r_\mathrm{th}$ and $z_\mathrm{th}$ are the given threshold parameters), then we mark it as an MM. Otherwise, it's labelled as an mM.

\section{RESULTS}\label{s4}
\subsection{\textit{J-M} Relation for Protohaloes}\label{s4s1}
We write the $J-M$ relation as
\begin{equation}\label{eqn25}
\frac{J}{J_0}=\beta\left(\frac{M}{M_0}\right)^\alpha,
\end{equation} 
where $J_0=10^{10}h^{-2}\mbox{M}_\odot \mbox{ kpc km s}^{-1}$ and $M_0=10^{10}h^{-1}\mbox{M}_\odot$. 

We find that in the $\Lambda$CDM model, at all redshifts, the $J-M$ relation for protohaloes (or haloes at $z=0$) can be well fitted as a power law (Figure \ref{f02}), with $\alpha (t)$ increasing from $\sim 1.5$ to $5/3$, and $\beta (t)$ evolving linearly with time in the beginning and reaching a constant finally (Figure \ref{f03}). This time-evolution behaviour can be understood using a three-regime scheme:

\begin{figure}
\includegraphics[width=245pt]{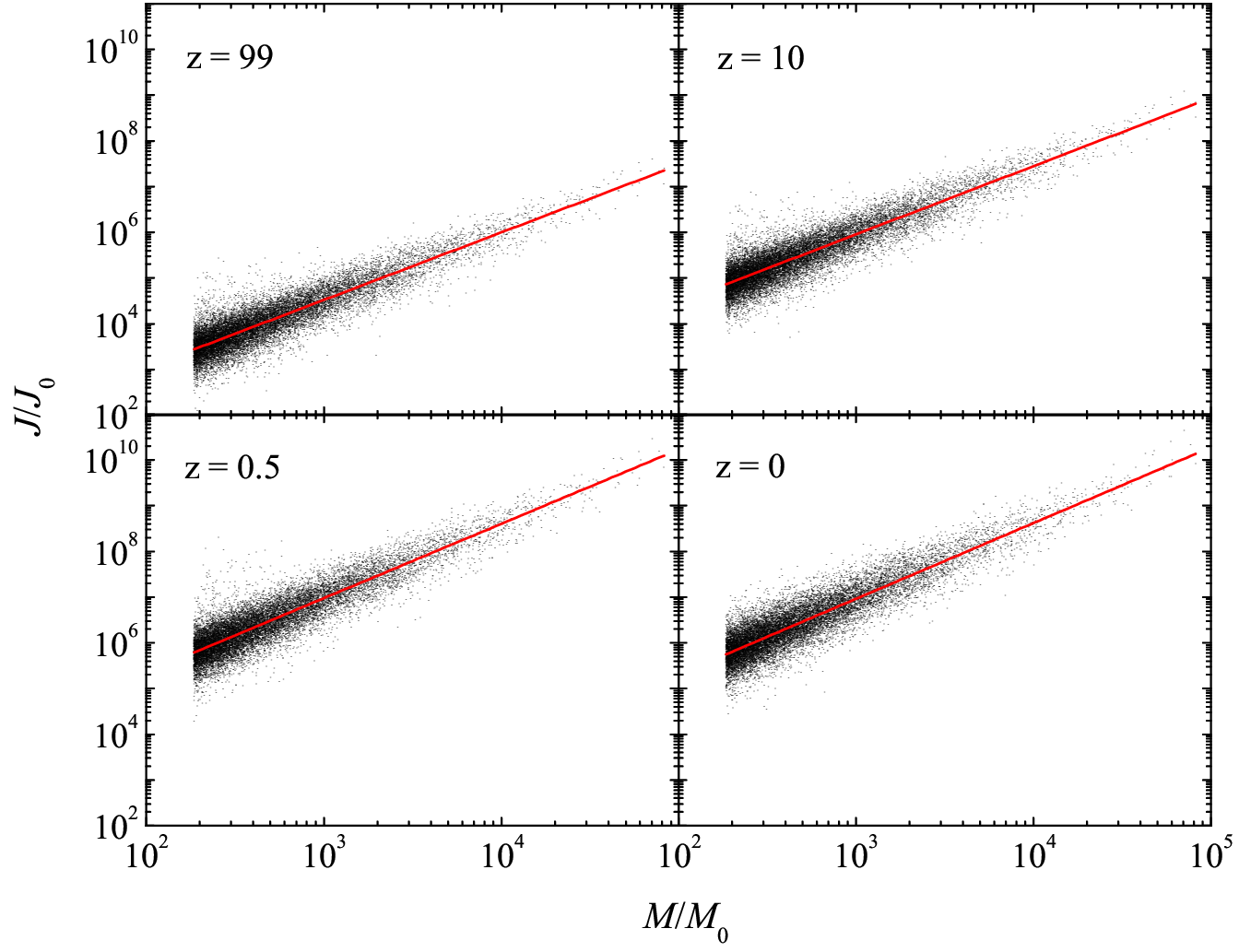}
\caption{$J-M$ relations for protohaloes at different redshifts in the $\Lambda$CDM model. The solid lines are best-fits for the $J-M$ relation using the APF fitting method. All of them can be well fitted as power laws, but with different $\alpha$ and $\beta$, as shown in Figure \ref{f03}. Similarly, we can observe such $J-M$ relations at every redshift in the AS-QCDM and scale-free models, but with different evolution behaviours of $\alpha$ and $\beta$ (see Figure \ref{f05}).}
\label{f02}
\end{figure}

\begin{figure}
\includegraphics[width=240pt]{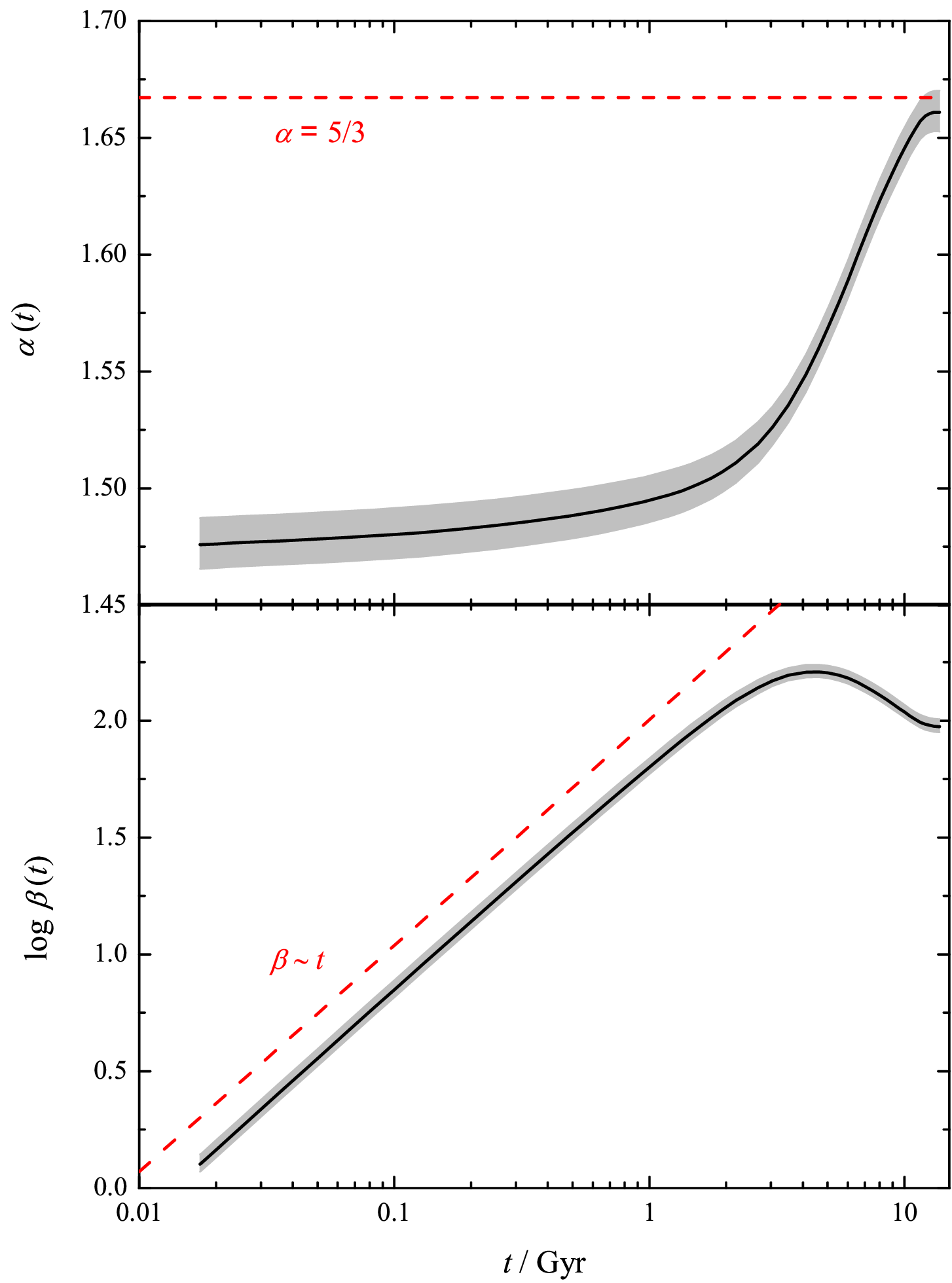}
\caption{Time evolution of the fitted $\alpha$ and $\log \beta$ in the $\Lambda$CDM model. The results are averaged over 10 realizations of $\Lambda$CDM512b simulations. The grey region shows the standard deviation among realizations.}
\label{f03}
\end{figure}

\begin{figure}
\includegraphics[width=240pt]{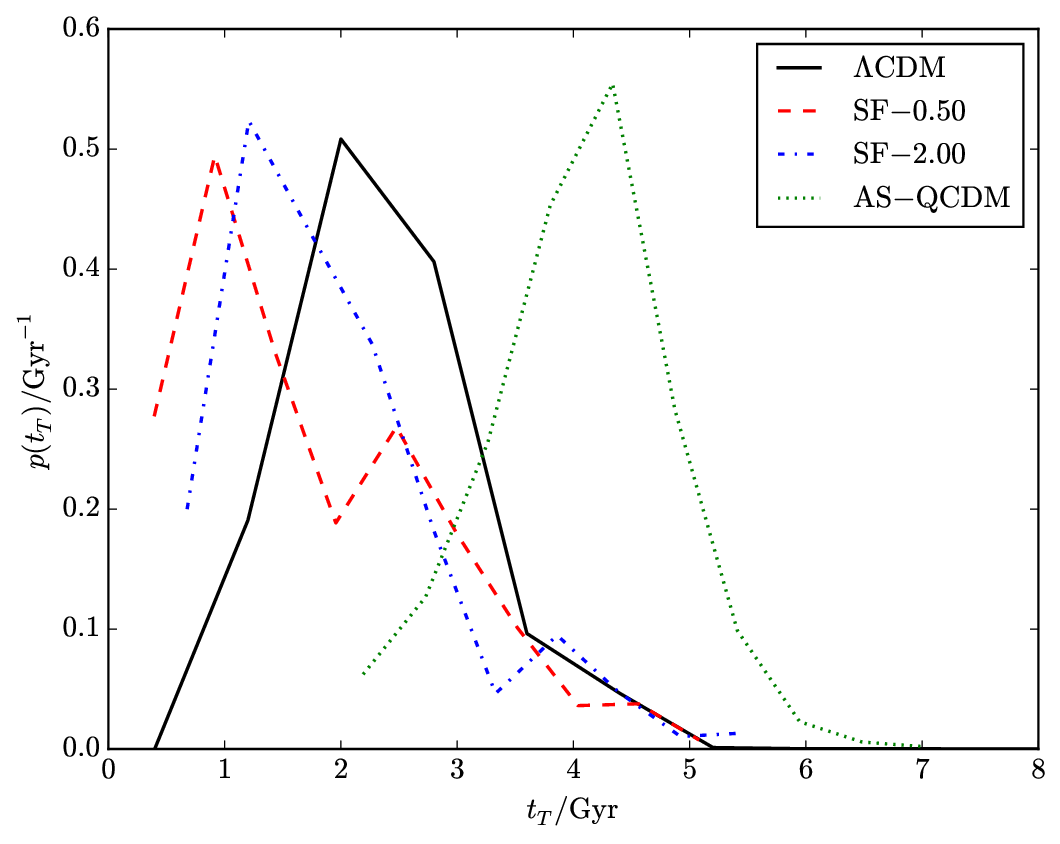}
\caption{Probability distribution of turnaround time $t_T$ for haloes in our simulations. The black solid, red dashed, blue dash-dotted and green dotted lines give the results from the $\Lambda$CDM, scale-free $n=-0.5$, $n=-2.0$ and AS-QCDM model respectively. To be clear, the results of $n=-1.0$, $n=-1.5$ scale-free models are not shown.}
\label{f04}
\end{figure}

(1) \textit{Linear regime}. In this stage, all protohaloes in our catalogue still evolve linearly. We adopt one of the methods in \citet{sugerman2000} to estimate the halo turnaround time $t_T$ as the earliest time that half of particles have negative radial physical velocity. The probability distribution of $t_T$ is plotted in Figure \ref{f04}. In our $\Lambda$CDM halo sample, almost all haloes reach turnaround during $t=1\sim 5$ Gyr. Only $0.3\%$ of haloes have turnaround time less than $t=1$ Gyr. As a result, we conservatively estimate the time period of the linear regime as $t<0.5$ Gyr (or $z>10$) for our $\Lambda$CDM halo sample. According to the discussion in Section \ref{s2s2}, in this regime, $\alpha$ remains constant and $\beta\propto t$. This is confirmed by our simulation results (Figure \ref{f03}). 

\begin{figure}
\includegraphics[width=240pt]{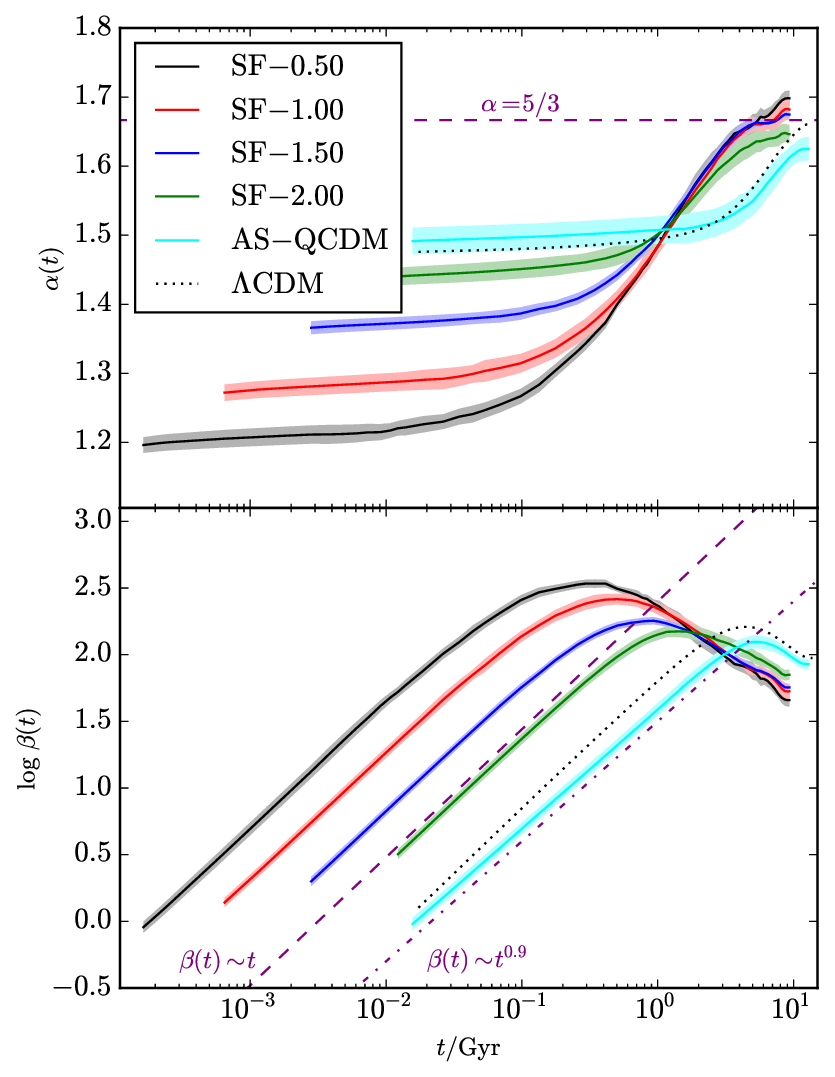}
\caption{Time evolution of $\alpha$ and $\beta$ in the AS-QCDM and scale-free models. The black, red, blue, green and cyan solid curves show the results from the SF-0.50, SF-1.00, SF-1.50, SF-2.00 and AS-QCDM simulations respectively. The shaded regions represent standard deviations among realizations. These evolution behaviours are similar to the $\Lambda$CDM case (dotted). In the linear regime, scale-free models' $\beta$ evolve approximately to $\beta \sim t$ (dashed line), while $\beta$ of the AS-QCDM model varies as $\beta \sim t^{0.9}$ (dash-dotted line).}
\label{f05}
\end{figure}

To help us understand the evolution behaviours of $\alpha(t)$ and $\beta(t)$, especially to test our predictions of Equation (\ref{eqn18}) and Equation (\ref{eqn19}) in the linear regime, we look at scale-free and AS quintessence dark energy (AS-QCDM, see Section \ref{s3s1}) simulations. Their $J-M$ relations are shown in Figure \ref{f05}.

\begin{figure}
\includegraphics[width=240pt]{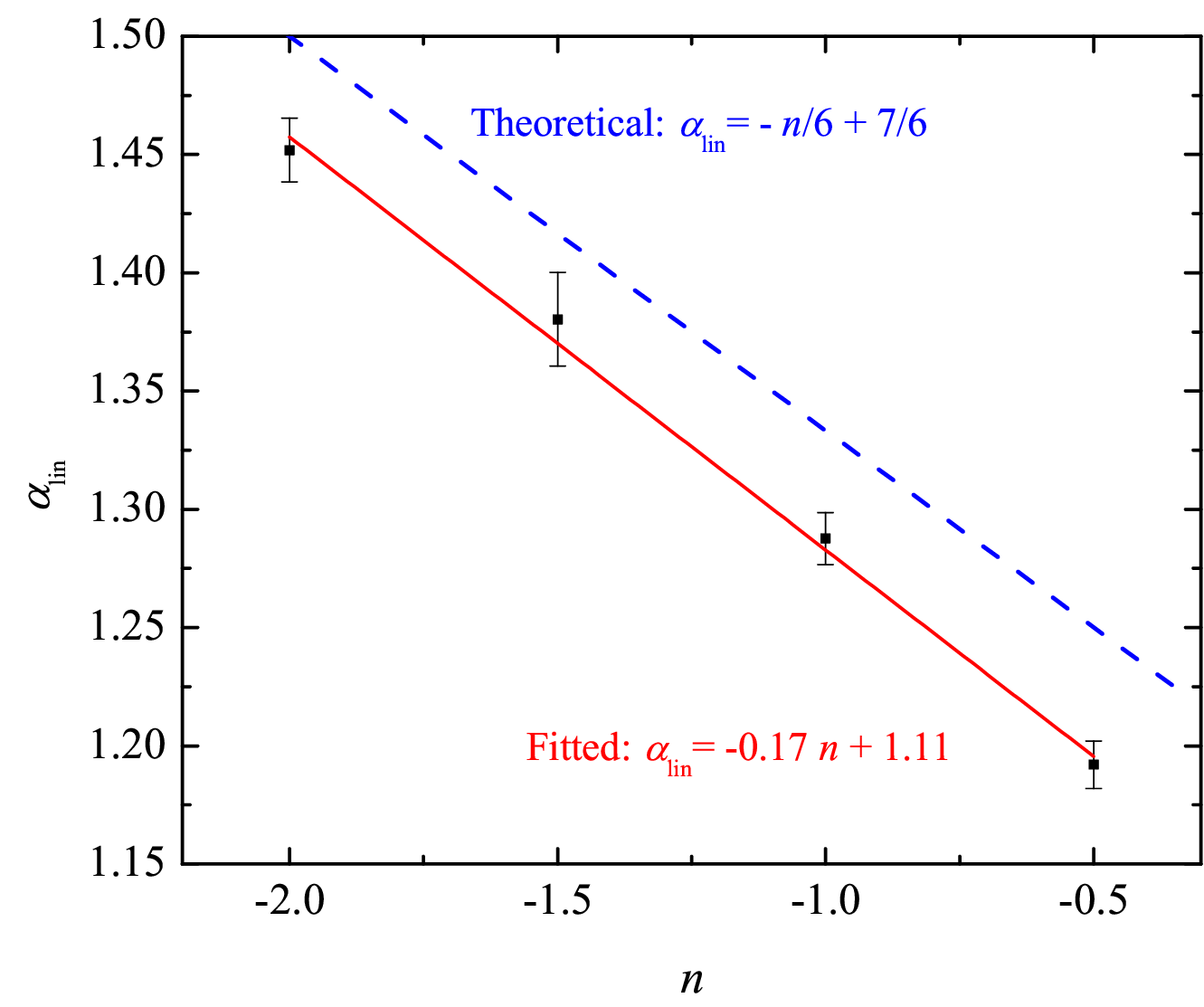}
\caption{$\alpha_\mathrm{lin}-n$ relation for scale-free models.}
\label{f06}
\end{figure}

Different models have different $\alpha$ in the linear regime, $\alpha_\mathrm{lin}$. For scale-free simulations, the $\alpha_\mathrm{lin}-n$ relation from simulations is shown in Figure \ref{f06}. It can be fitted as $\alpha_\mathrm{lin}=-0.17n+1.11$, which has a deviation of $\sim 0.05$ in the $y$-intercept from the theoretical prediction $\alpha_\mathrm{lin}=-n/6+7/6$ [Equation (\ref{eqn18})]. This comes from the underlying moment of inertia-mass relation ($I-M$ relation). We have used $I\propto M^{5/3}$ when deriving Equation (\ref{eqn18}) assuming protohaloes have similar shapes. But the simulated protohaloes follow a slightly different relation, $I\propto M^{1.56\pm 0.01}$, since their triaxial ratios are usually not perfectly similar. After taking into account such effect, the numerical results agree with our prediction. 

We can use the effective index $n_\mathrm{eff}(k)= d\ln P(k)/d\ln k$ to understand the value of $\alpha_\mathrm{lin}$ in the $\Lambda$CDM and AS-QCDM model. For protohalo scale ($\sim 1h^{-1}\mbox{Mpc}$) in such models, $n_\mathrm{eff} \sim -2.0$ and Equation (\ref{eqn18}) gives $\alpha_\mathrm{lin} \sim 1.5$.

The linear regimes span different periods in different models (Figure \ref{f05}). This can be understood by looking at the halo turnaround time $t_T$ in different models, as shown in Figure \ref{f04}. A scale-free model with less negative $n$ has more power in small scale perturbations and thus make haloes turn around earlier. Although we start with the same power spectrum in the $\Lambda$CDM and AS-QCDM model, haloes in the AS-QCDM model tend to have larger $t_T$, since the AS-QCDM model contains a larger fraction of dark energy and thus a faster expansion rate at high redshifts, consequently delaying the halo turnaround time.

\begin{figure}
\includegraphics[width=240pt]{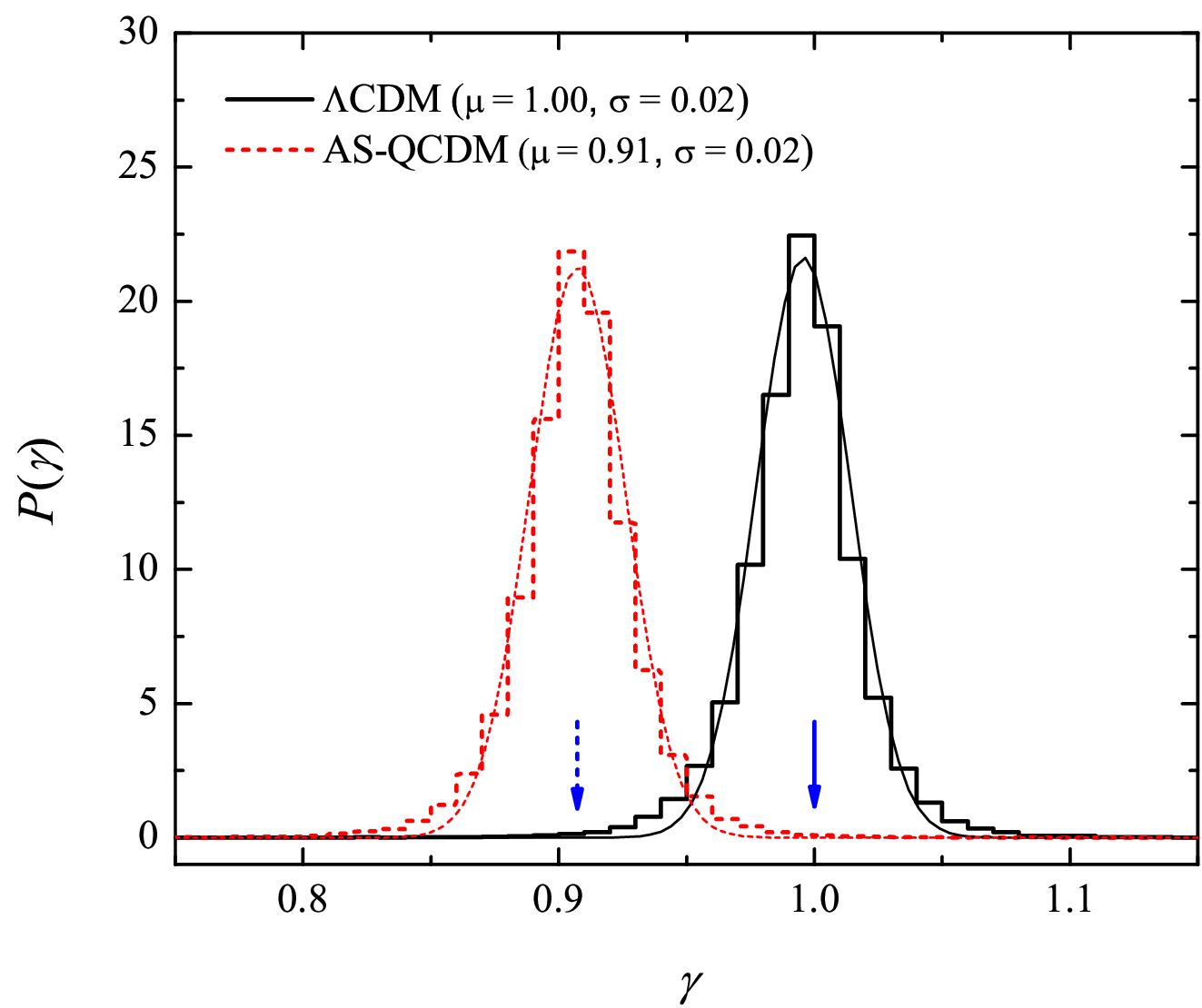}
\caption{Probability distribution function (PDF) of the fitted $\gamma$ for $\Lambda$CDM (thick solid) and AS-QCDM (thick dashed) model, fitted as a Gaussian PDF $p(\gamma)=(1/\sigma\sqrt{2\pi})\exp[-(\gamma-\mu)^2/2\sigma^2]$ shown in the thin solid and dashed lines respectively. The best-fitted $(\mu, \sigma)$ are $(1.00, 0.02)$ and $(0.91, 0.02)$ for $\Lambda$CDM and AS-QCDM model respectively. Numerically calculated TTT predictions, $\gamma = 2x+y-1$, are marked with arrows.}
\label{f07}
\end{figure}

In the linear regime, scale-free models follow a similar $\beta \sim t$ as the $\Lambda$CDM model, since both of them are matter-dominated at high redshifts. However, the AS-QCDM model has a significantly different growth factor $D(t)$ from $\Lambda$CDM in the linear regime. As we can see from the lower panel of Figure \ref{f05}, in the linear regime, $\beta_{\mathrm{AS-QCDM}} \sim t^{0.9}$. This is consistent with the numerically calculated $J \sim a^2\dot{D} \sim t^{0.9132}$, and thus supports Equation (\ref{eqn19}) (see Table \ref{t03}). We can also look at the time evolution of each halo's angular momentum, which has the same dependence on $a^2\dot{D}$ according to TTT. Assuming $J(t)\sim t^\gamma$, we fit $\gamma$ for each protohalo in its linear regime and obtain a Gaussian probability distribution $p(\gamma)$ for all haloes, shown in Figure \ref{f07}. The mean of $\gamma$ in each cosmological model agrees with the TTT prediction as expected.

Since the time-evolution behaviours of $J-M$ relations in the AS-QCDM and scale-free models are qualitatively similar to that of $\Lambda$CDM (Figure \ref{f05}), in the following discussion, we will mainly present the $\Lambda$CDM results. Similar arguments and explanations can be applied to the AS-QCDM and scale-free models.

\begin{table*}
 \centering
  \caption{Comparison between the numerical calculations and protohalo fitting results.}
  \begin{tabular} {@{}ccccc@{}}
  \hline
   Model & $x$ & $y$ & $2x+y-1$ & $\gamma$ Fitted From Protohaloes\\
         & $(a\sim t^x)$ & $(D\sim t^y)$ & $(J\sim t^{2x+y-1})$ & $(J\sim t^\gamma)$\\
 \hline
   $\Lambda$CDM & 0.6667 & 0.6667 & 1.0001 & 1.00$\pm$0.02\\
   AS-QCDM & 0.6684 & 0.5764 & 0.9132 & 0.91$\pm$0.02\\
\hline
\end{tabular}
\label{t03}
\end{table*}

(2) \textit{Non-linear regime}. After the linear regime, some protohaloes (especially the small mass ones) start to evolve nonlinearly. For our $\Lambda$CDM halo catalogue, this regime ranges from $t=0.5$ Gyr to present. 

In this regime, $\alpha$ increases monotonically while $\beta$ reaches a maximum and decreases a little. Notice that even when almost all haloes have reached turnaround (e.g. in Figure \ref{f04}, $99.95\%$ of $\Lambda$CDM haloes reached turnaround after $t=6$ Gyr), the $J-M$ relation still evolves. This is different from TTT's prediction. We conclude that nonlinear effects play an important role in the time evolution of the $J-M$ relation.

The evolution of $\beta$ is similar to that of a halo's angular momentum \citep[][]{sugerman2000,porciani2002a}. The decrease of a halo's angular momentum, or $\beta$ in the $J-M$ relation, is due to its nonlinear interactions with the surrounding matter which lead to the redistribution of angular momenta. 

(3) \textit{Virial regime}. Once the haloes become virialized and if they experience no merger events, their angular momenta stop evolving, and thus the $J-M$ relation becomes stable, with $\alpha$ and $\beta$ both becoming constants. In particular, $\alpha$ approaches $5/3$, which can be explained using the mechanical equilibrium argument. 

To quantify the virialization of haloes at $z=0$, we use the offset parameter defined as

\begin{equation}\label{eqn26}
s=\frac{|{\bf r}_\mathrm{mb}-{\bf r}_\mathrm{cm}|}{R_\mathrm{vir}},
\end{equation}
where ${\bf r}_\mathrm{mb}$, ${\bf r}_\mathrm{cm}$ and $R_\mathrm{vir}$ are the position of the most bound particle within a halo, center-of-mass of a halo and halo's virial radius respectively. Relaxed haloes have small $s$; haloes having $s<0.1$ are usually regarded as relaxed \citep[e.g.][]{donghia2007}. In our $z=0$ halo sample ($\Lambda$CDM512b simulation), $\log s$ distributes normally with a mean of $-1.12$ and standard deviation of $0.28$.

As a complementary way to quantify the relaxation of haloes, we also calculate the virial parameter
\begin{equation}\label{eqn27}
\eta=\frac{2K}{|U|},
\end{equation}
where $K=\sum_i m_i v^2_i/2$ and $U=\sum_{i=0}^{N-1} \sum_{j=i+1}^N-\frac{Gm_i m_j}{r_{ij}}$ are the halo's kinetic and potential energy.  According to the virial theorem, $\eta$ becomes 1 when an isolated object relaxes. For our halo catalogue at $z=0$, the mean (median) value of $\eta$ is $1.11$ $(1.08)$, with a standard deviation of $0.16$. The distribution of $s$ and $\eta$ for our halo sample indicates that most haloes are close to being virialized at $z=0$.

To see more explicitly the correlation between $\alpha=5/3$ and virialization, we divide the haloes at $z=0$ into two subsets: $s\leq 0.1$ and $s>0.1$ and fit the $J-M$ relation for them separately. The best-fits are $\alpha=1.65\pm 0.01, \log \beta = 1.82\pm 0.03$ for $s\leq 0.1$ haloes and $\alpha=1.75\pm 0.02, \log \beta = 1.74\pm 0.04$ for $s>0.1$ haloes. The threshold value of $0.1$ here is not special. Changing this threshold value for $s$ does not change the conclusion that $\alpha$ becomes $5/3$ for virialized haloes, but is significantly different from $5/3$ for non-virialized ones. 

\begin{figure*}
\includegraphics[width=500pt]{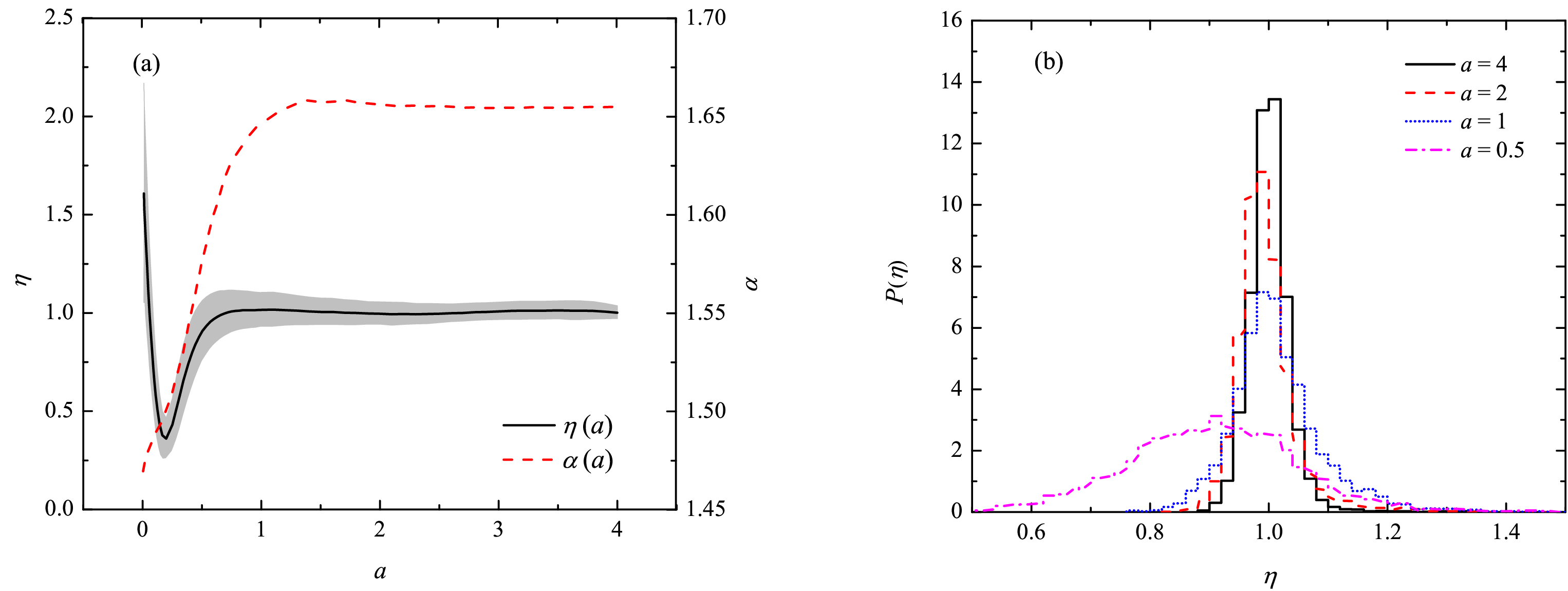}
\caption{(a) Time evolution of $\alpha$ (dashed) and $\eta$ (solid) for haloes identified at $a=4$. The solid line represents the mean value of $\eta$ for the whole halo sample and the grey region shows the $\pm 1\sigma$ errors. (b) Probability distribution of $\eta$ at different scale factors: $a=0.5$ (dash-dotted), $a=1$ (dotted), $a=2$ (dashed) and $a=4$ (solid).}
\label{f08}
\end{figure*}

In addition, we plot in Figure \ref{f08} the time evolution of the virial parameter. Especially, to illustrate the correlation between the evolution of $\alpha$ and $\eta$ more clearly, we use future haloes identified at $a=4$ (or $z=-0.75, t=35.65$ Gyr) since the majority of them will be fully virialized. For $a=4$ haloes, we trace the particles back and perform the same fitting for the $J - M$ relation, as for haloes identified at $a=1$. As shown in Figure \ref{f08}(a), when most haloes become virialized, that is, the mean $\bar{\eta} \sim 1$ and standard deviation $\sigma$ becomes small enough, $\alpha$ reaches a stable value $\sim 5/3$. Notice that the mean $\bar{\eta}$ reaches $1$ at $a\sim 0.8$, but $\alpha$ is still varying at this moment. This is due to the fact that there are still some haloes that are far from virialization, as shown by the relatively large standard deviation $\sigma_\eta$. For example, at $a=1$, $\sigma_\eta/\bar{\eta} = 10.0\%$, while at $a=4$, $\sigma_\eta/\bar{\eta} = 3.5\%$. As time evolves, $\bar{\eta}$ gets closer to $1$ and the dispersion becomes smaller [Figure \ref{f08}(b)].

With this three-regime scheme, we can understand the observed time evolution of the $J-M$ relation from N-body simulations. Especially, we show clearly that the observed exponent $\alpha=5/3$ correlate with virialization. On the other hand, TTT is able to explain the $J-M$ relation in the linear regime if we consider the effects from smoothing the potential term. In the linear regime, $\alpha$ depends on the power index of the power spectrum, whereas the time evolution of $\beta$ contains the information of the underlying cosmological model. The three-regime scheme can also be used to understand the $J-M$ relations for haloes in different environments and with different merging histories, as we will discuss in Sections \ref{s4s2} and \ref{s4s3}.

\subsection{Dependence on Environments}\label{s4s2}

\begin{figure}
\includegraphics[width=240pt]{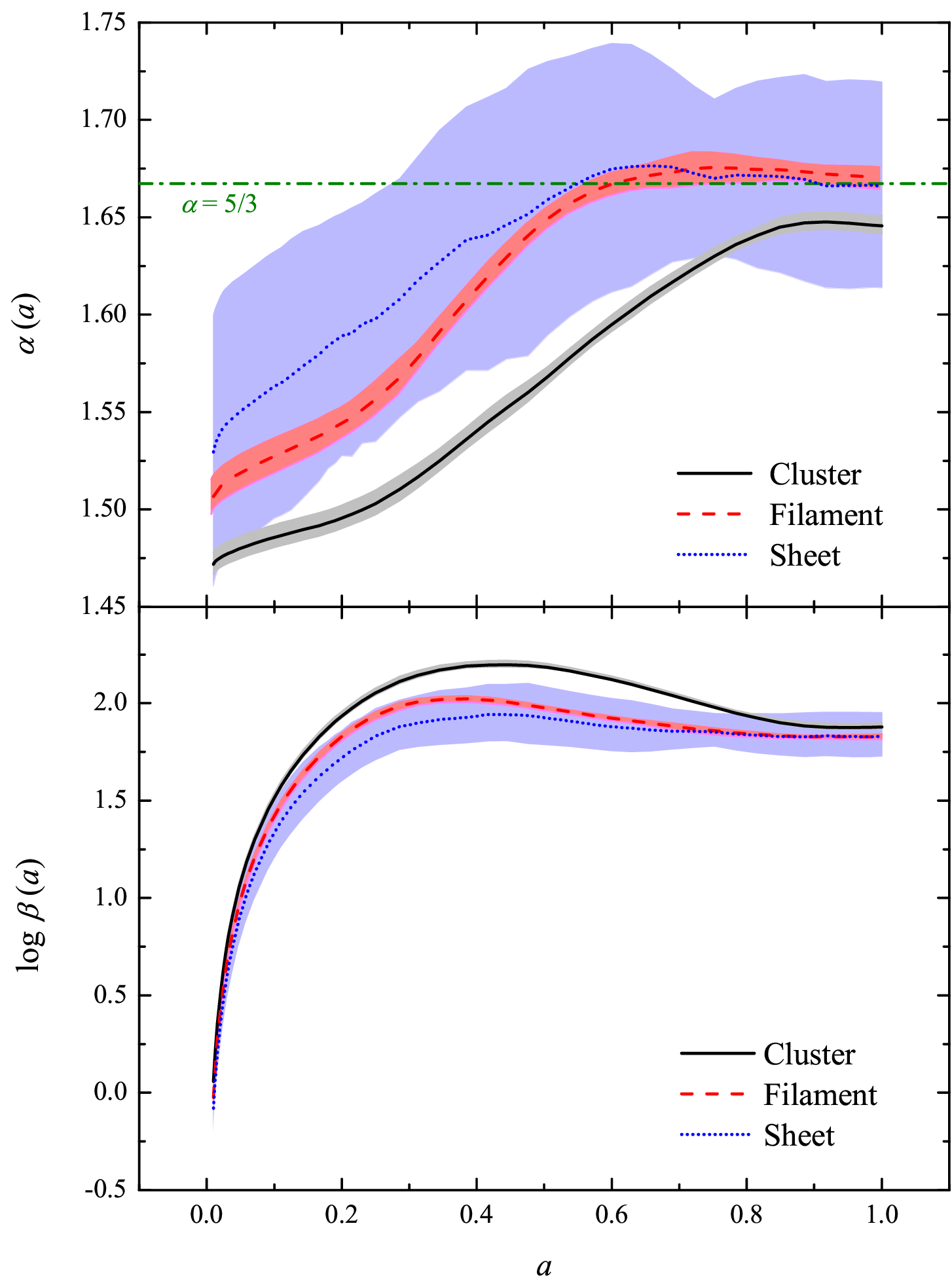}
\caption{Time evolution of $\alpha$ and $\beta$ for haloes in different environments in the $\Lambda$CDM model. Solid, dashed and dotted curves show the results for cluster, filament and sheet haloes respectively. The shaded regions are standard deviations among realizations. There are few void haloes in our simulation and thus they are not included.}
\label{f09}
\end{figure}

The time evolution of the $J-M$ relations for haloes in clusters, filaments and sheets are shown in Figure \ref{f09}. In our $\Lambda$CDM simulations, there are too few void haloes to perform a reliable fit for the $J-M$ relation, and thus we do not discuss them here. The numbers (fractions) of cluster, filament, sheet and void haloes at $z=0$ are $\sim 14000$  $(45.15\%)$, $\sim 16000$ $(51.60\%)$, $\sim 1000$ $(3.22\%)$ and $\sim 10$ $(0.03 \%)$ respectively in a $\Lambda$CDM512b simulation with at least $200$ halo particles ($M \geq 9.2 \times 10^{11}h^{-1}$M$_\odot$). Note that we only perform environmental classifications on the haloes at $z=0$ and not as a function of redshift.

From Figure \ref{f09}, we can see that filament and sheet haloes have a larger $\alpha$ in the linear regime. This is due to the deviation of the power index from $5/3$ in the underlying $I-M$ relation (Table \ref{t04}). Cluster haloes experience more nonlinear effects and their protohaloes usually have more complicated and non-similar shapes. Their $I-M$ relation deviates more from a power index of $5/3$, which leads to a larger deviation of $\alpha$ from $7/6-n_\mathrm{eff}/6$ for their $J-M$ relation. 

\begin{table*}
 \centering
  \caption{The power index $w$ of the $I-M$ relation, $I\propto M^w$, for haloes in different environments.}
  \begin{tabular} {@{}cccc@{}}
  \hline
   Type & Cluster & Filament & Sheet \\
   $w$ & $1.54\pm 0.01$ & $1.58\pm 0.01$ & $1.58\pm 0.02$\\
\hline
\end{tabular}
\label{t04}
\end{table*}

In addition, filament and sheet haloes' $J-M$ relations become stable earlier than cluster haloes. For example, the filament haloes' $\alpha$ stabilizes to a value near $5/3$ at $a\approx 0.7$, while the $\alpha-a$ curve for cluster haloes reaches a plateau at $a\approx 0.9$. A similar behaviour can be observed for $\log \beta$. This is due to the fact that filament and sheet haloes tend to locate in relatively low density regions \citep[see e.g.][]{hahn2007}, experience less nonlinear effects, and enter the equilibrium regime earlier. But still haloes of each classified type can span a wide range of densities, and this is likely the cause for the large scatters of sheet haloes, which have relatively small number, in Figure \ref{f09}. Note that we only study the case of $\lambda_\mathrm{th}=0$ here. A different $\lambda_\mathrm{th}$ can lead to different fractions of classified types, as shown by \citet[]{forero2009}.

The environmental dependences of $J-M$ relations in the AS-QCDM and scale-free models are qualitatively similar to the $\Lambda$CDM case.

Thus, by dividing the haloes into different environments, we can see clearly how the nonlinear effects affect the $J-M$ relation. We have also shown that the differences of $J-M$ relations in different environments can be explained using the three-regime scheme.

\subsection{Dependence on Merging Histories}\label{s4s3}

\begin{figure}
\includegraphics[width=240pt]{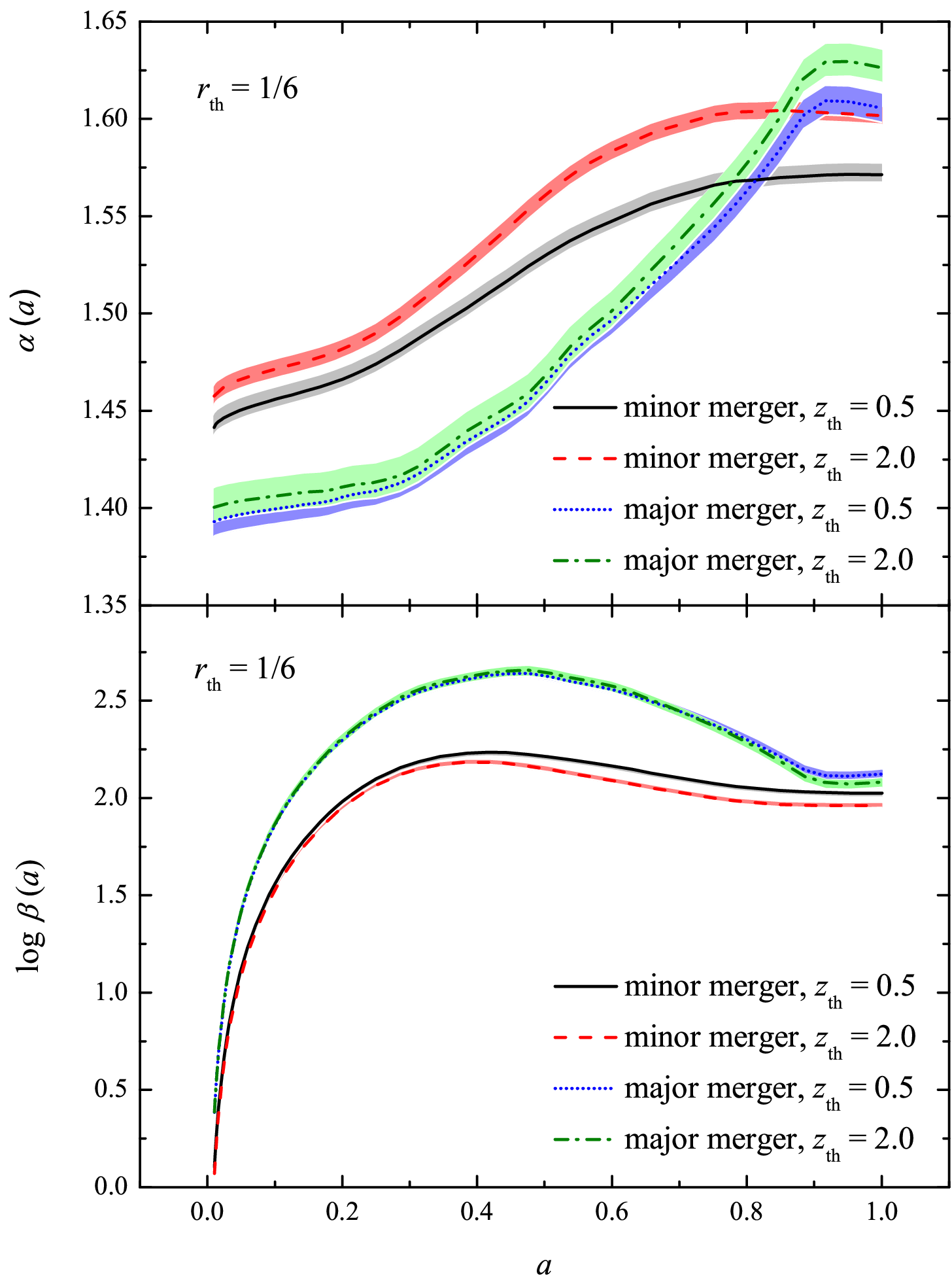}
\caption{Similar to Figure \ref{f09}, but here we show the time evolution of $J-M$ relations for minor merger (mM) and major merger (MM) haloes. To be clear, we only plot the results with $r_\mathrm{th}=1/6$ and $z_\mathrm{th}=0.5, 2.0$. Cases for other $r_\mathrm{th}$ and $z_\mathrm{th}$ are similar.}
\label{f10}
\end{figure}

We study the $J-M$ relation for mM and MM haloes with different threshold parameters $r_\mathrm{th}=1/6, 1/5, 1/3$ and $z_\mathrm{th}=0.5, 1.0, 2.0$. The results are shown in Figure \ref{f10}. To be simple and clear, we only plot the cases of $r_\mathrm{th}=1/6$ and $z_\mathrm{th}=0.5, 2.0$. Other cases lead to similar conclusions.

$\alpha$ has a larger initial value for the mM halo subset and becomes stable earlier compared to the MM halo subset. For both types of haloes, $\alpha$ tends to be larger for higher $z_\mathrm{th}$. These can be understood as following: (1) MM haloes usually have more complicated protoshapes and thus larger deviations from $5/3$ for the power index of the $I-M$ relation. This leads to a larger deviation for $\alpha_\mathrm{lin}$ from $7/6-n_\mathrm{eff}/6$. (2) The $\alpha$ and $\log \beta$ for mM haloes become stable earlier because they go through less nonlinear evolution. (3) By increasing $z_\mathrm{th}$, we exclude haloes with more complicated evolution in the mM subset, and thus the final $\alpha$ has a value closer to $5/3$. (4) $\alpha_f$, the final values of $\alpha$, for both mM and MM subsets are smaller than $5/3$. For example, with $z_\mathrm{th}=2.0$, $\alpha_f=1.60\pm 0.01$ for mM haloes and $\alpha_f=1.63\pm 0.01$ for MM haloes. This is due to the fact that unrelaxed haloes have larger effects on these subsets. If we exclude haloes with offset parameter $s>0.1$, $\alpha_f$ for mM and MM increase to $1.62\pm 0.01$ and $1.64\pm 0.02$, which are closer to $5/3$. 

\section{SUMMARY AND DISCUSSIONS}\label{s5}
We have used N-body simulations to study the time-evolution of the $J-M$ relation. From our results emerges a picture of the origin and evolution of the $J-M$ relation in the $\Lambda$CDM model:

At high redshifts, when all haloes in our sample still evolve linearly, $\alpha$ is a constant of $\sim 1.5$ and $\beta$ increases linearly with time. We show that this can be explained using the tidal torque theory if we carefully consider the mass dependence introduced by the smoothing of potential field [Equation (\ref{eqn18}) and Equation (\ref{eqn19})], needed for keeping the validity of the Zel'dovich approximation and Taylor approximation up to second order. In the nonlinear regime, $\alpha$ increases monotonically and $\beta$ gradually reaches a maximum and decreases. Finally, in the virial regime when the majority of haloes become virialized, $\alpha$ becomes a constant $5/3$ and $\beta$ stabilizes.

This time evolution picture enables us to discriminate among possible explanations. We show that the empirically observed $\alpha=5/3$ is consistent with the mechanical equilibrium of haloes. On the other hand, TTT successfully explains the $J-M$ relation in the linear regime. Haloes in different environments and with different merging histories show different time evolution behaviours of the $J-M$ relation. The nonlinear effects drive the $J-M$ relation in the linear regime to the one we observed.

\citet[]{antonuccio2010} also looked at the evolution of the $J-M$ relation and found that $\alpha$ is compatible to $5/3$ at high redshift but becomes slightly smaller than $5/3$ recently (see their Figure 2). However, one should notice that their $J-M$ relations are fitted from halo samples identified at different redshifts, which are different from ours from the trace-back picture.

The three-regime scheme implies that for different cosmologies, in the linear regime, $\alpha$ has different values according to the initial power spectrum and protohaloes' shapes, and $\beta$ evolves with different rates depending on the scale factor and growth rate. Thus, in the linear stage, the $J-M$ relation is quite sensitive to the underlying cosmological model. In the nonlinear regime, the evolution of the $J-M$ relation depends on the details of nonlinear collapse, mergers and other nonlinear effects in a cosmology. In a cosmological model with more haloes in the denser environment and experiencing major mergers, the $J-M$ relation will take more time to reach the stable state. When all haloes become virialized and go through no merger events, the corresponding $J-M$ relation stabilizes. $\alpha$ will lose the memory of the initial power spectrum and background cosmology, and has a universal value of $5/3$. Whether $\beta$ in the virial stage depends on the initial power spectrum and background cosmology is an interesting question. There is no exact analytical theory to calculate the final spin of a halo. It is shown that the spin parameters $\lambda$ \citep[][]{peebles1971, bullock2001} for virialized haloes have no substantial dependence on the initial conditions and background cosmology \citep[see e.g. ][, etc.]{barnes1987,bullock2001,maccio2008,carlesi2012} Here, in the virial regime, the dependence of $\beta$ on the background cosmology is also weak (see Figure \ref{f05}). Recently, \citet{lee2013} showed that modified gravity could spin up galactic haloes with $M\leq 10^{11}h^{-1}M_\odot$. It will be interesting to see whether modified gravity has great effects on the virial $J-M$ relation.

Although our simulations are only for dark matter particles, some behaviours of the $J-M$ relation we found can be generalized to baryonic matter. For example, in the linear regime, we expect that $\alpha$ for baryonic matter also depends on its initial power spectrum and the protogalaxy's shapes. $\beta$ also increases proportionally to $a^2\dot{D}$. In the virial regime, once galaxies become virialized, their $J-M$ relation remains unchanged. $\alpha$ equals to $5/3$ because it's a result of virialization. Indeed, the spin parameters for baryonic and dark components have been shown to correlate in cosmological hydrodynamic simulations \citep[][]{bosch2002,chen2003,sharma2005,gottlober2007,kimm2011}. However, in the nonlinear regime, due to the diverse baryonic physics, such as radiative cooling, star formation, supernovae and AGN feedback, etc., the evolution of galaxies' spins, disks, spin alignments and other properties is quite complicated \citep[see e.g.][]{bailinetal2005,bailinsteinmetz2005,libeskind2007,schafer2009,roskar2010,schewtschenko2011},and protogalaxies' $J-M$ relation in this stage might have different behaviours from the dark matter and needs further investigation. The baryonic processes in the nonlinear regime are also key elements to explain the observed offset between the spirals and ellipticals' $J-M$ relations, as discussed in \citet{romanowsky2012}.

Another question related to the baryon physics is how the angular momentum transfer between dark and baryonic matter affects our results. It has been shown that from the hydrodynamical simulations, the baryonic physics mainly spins up the inner part of a halo, and has minor effects on the whole halo's spin \citep[e.g. ][]{bett2010,bryan2013}. Thus, we also expect that our results about the dark matter haloes' $J-M$ relation should not change significantly when one adds baryon physics into the simulations.

Although it was discovered in 1960s, the $J-M$ relation is still an ongoing research topic in observations \citep[e.g.][]{romanowsky2012,fall2013}, and a complete theoretical explanation is needed. Here, we give an updated picture of this relation for the dark matter part.

\acknowledgments
The authors thank Yipeng Jing and Donghai Zhao for useful suggestions and Weipeng Lin for helps in performing the simulations. The simulations in this article are performed in the Chinese University Information Technology Services Centre clusters and the supercomputing platform of Shanghai Astronomical Observatory. Jiayu Tang is partially supported by a postdoctoral fellowship by the Chinese University of Hong Kong. The work is supported by Chinese University of Hong Kong Direct Grants 4053013 and 4053069.

\appendix

\section{Numerical box size studies}\label{as1}
Previous studies \citep[e.g.][]{bagla2005,power2006} showed that small numerical box size would reduce the number of massive haloes and lower the haloes' spin parameters. Therefore, we expect that the numerical box size should affect the $J-M$ relation. To study its effects and find out a suitable box size, we performed several N-body simulations using the method of \citet{power2006}. In this method, we chopped the long wavelength perturbations in different degrees to mimic different box sizes $L_\mathrm{chop}$. The smallest wave vector $k_\mathrm{min}$ depends on the chopping factor $f_\mathrm{chop}$ as
\begin{equation}\label{aeqn01}
k_\mathrm{min}=\frac{2\pi}{L_\mathrm{chop}}=\frac{2\pi}{f_\mathrm{chop}L_\mathrm{box}}.
\end{equation}
The initial conditions with different $f_\mathrm{chop}$ were generated with the same random seed so that we could compare the simulations directly. The simulation details are listed as following: $\Omega_m = 0.28, \Omega_\Lambda=0.72, \Omega_bh^2=0.024, h=0.7, N_p=512^3, L_\mathrm{box}=200 h^{-1}$Mpc, $\sigma_8=0.8, n_s=0.96$ and comoving softening length $\epsilon = 20.0 h^{-1}$kpc. We performed simulations with chopping factors $f_\mathrm{chop}=1.0, 0.75, 0.5, 0.25, 0.1$.

\begin{figure}
\centering
\includegraphics[width=240pt]{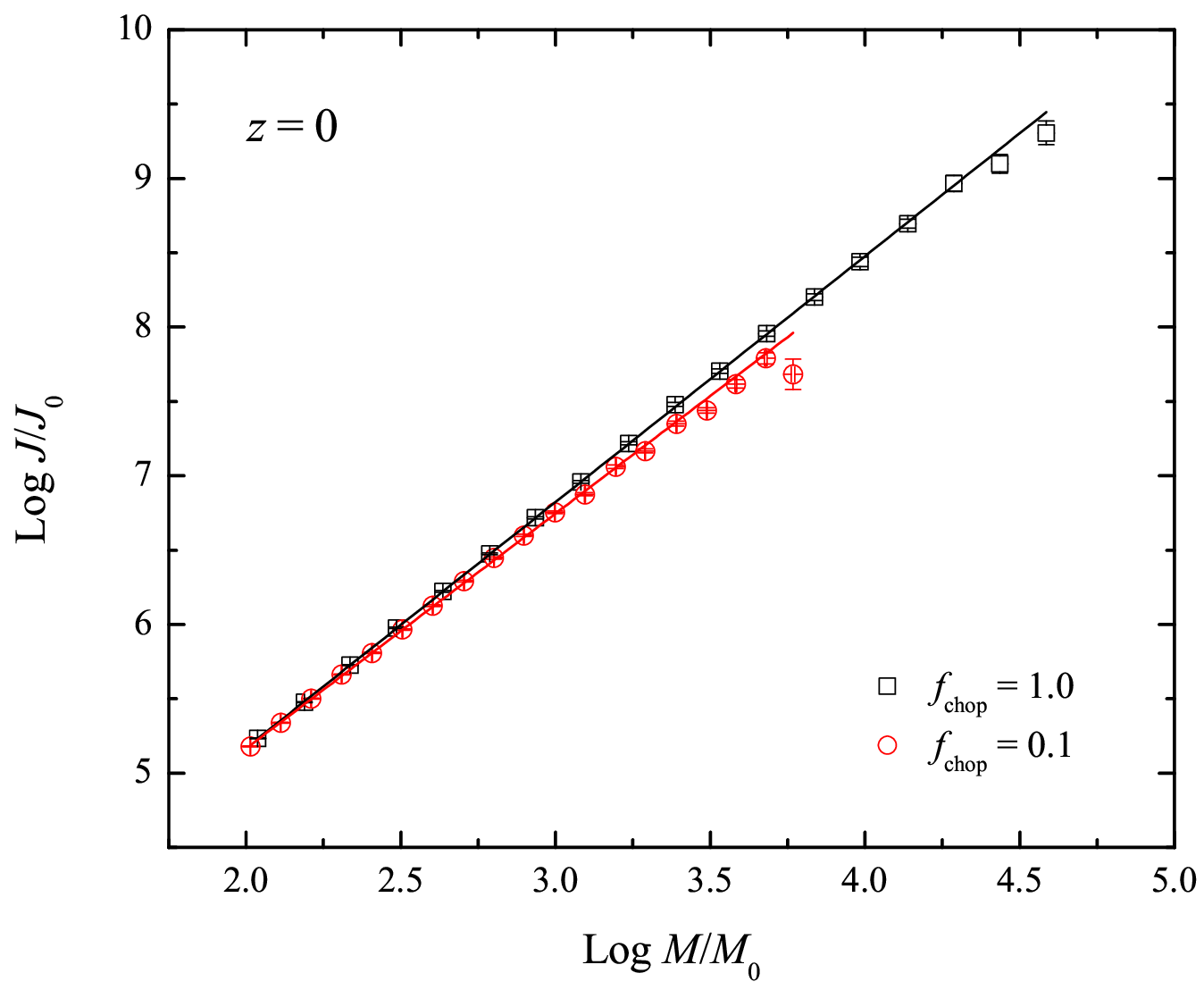}
\caption{$J-M$ relations for different simulations at $z=0$. For clarity, we only plot results for $f_\mathrm{chop}=1.0$ (square) and $0.1$ (circle), the two extreme cases. Here we have divided the haloes into several bins according to their masses. The points and error bars in the plot show us the means and standard deviations of $\log J$ and $\log M$ in different bins.}
\label{f11}
\end{figure}

Figure \ref{f11} presents the $J-M$ relations at $z=0$ for simulations with $f_\mathrm{chop}=1.0$ and $0.1$. The angular momenta are lower in simulations with smaller $f_\mathrm{chop}$. This is especially obvious for massive haloes. For the simulation with $f_\mathrm{chop}=0.1$, there is no halo with mass $M>10^{14}h^{-1}$M$_\odot$. The best-fitted $J-M$ relations are shown in Table \ref{t05}. We find that simulations with $f_\mathrm{chop}=0.5$ and $0.75$ have convergent results to the no-chopping simulation. Therefore, to study the $J-M$ relation, $L_\mathrm{box}$ should be at least $100 h^{-1}$Mpc.

\begin{table*}
 \centering
  \caption{Best-fits of the $J-M$ relation for different $f_\mathrm{chop}$ simulations.}
  \begin{tabular} {@{}ccc@{}}
  \hline
   $f_\mathrm{chop}$ & $\alpha$ & $\log\beta$ \\
 \hline
   1.0  & $1.654\pm 0.004$ & $1.863\pm 0.010$ \\
   0.75 & $1.649\pm 0.004$ & $1.874\pm 0.010$ \\
   0.5  & $1.652\pm 0.004$ & $1.867\pm 0.010$ \\
   0.25 & $1.628\pm 0.004$ & $1.918\pm 0.010$ \\
   0.1  & $1.582\pm 0.004$ & $2.002\pm 0.010$ \\
 \hline         
\end{tabular}
\label{t05}
\end{table*}

\section{Halo resolution studies}\label{as3}
What resolution is needed for studying haloes' angular momenta? Or, what is the suitable minimum particle number $N_\mathrm{min}$ for a halo to give converged angular momentum? To our knowledge, there is a wide range for $N_\mathrm{min}$ (from $\sim 50$ to $\sim 1000$) used in the literature. Here, to obtain a better estimation of $N_\mathrm{min}$, we performed several simulations with different resolutions. All parameters are the same for these simulations, except for the particle number $N_p$. In our simulations, $N_p = 512^3, 256^3, 128^3$ and $64^3$. Other parameters are: $\Omega_m = 0.28, \Omega_\Lambda=0.72, \Omega_bh^2=0.024, h=0.7, L_\mathrm{box}=100 h^{-1}$Mpc, $\sigma_8=0.8, n_s=0.96$ and comoving softening length $\epsilon = 10.0 h^{-1}$kpc. The same scale perturbations in all of these simulations have same phases, and thus it allows us to compare the haloes by one-on-one mapping.

We firstly identified haloes using AHF in all simulations with $\Delta_\mathrm{vir}=98$ and $N_\mathrm{min}=10$, and then we mapped the haloes in $N_p = 64^3, 128^3, 256^3$ simulations to haloes in the $N_p=512^3$ simulation by requiring that each corresponding halo pair has similar locations and masses. Due to having less particles to sample the density field in low resolution simulations and the noise from the halo finder, the haloes in low resolution simulations usually don't have perfectly identical positions and masses as those in high resolution simulations. Thus, it's a nontrivial task to map the haloes. 

We define two parameters related to position and mass differences as
\begin{equation}\label{aeqn02}
\tau_\mathrm{pos}=\frac{\sqrt{(x_n-x_{512})^2 + (y_n-y_{512})^2 + (z_n-z_{512})^2}}{R_{512}},
\end{equation}
\begin{equation}\label{aeqn03}
\tau_\mathrm{mass}=\frac{|M_n-M_{512}|}{M_{512}},
\end{equation}
where $x_n, y_n, z_n, R_n, M_n$ are the $x-,y-,z-$ positions, virial radius and mass of a halo in the $N_p=n^3$ simulation. We firstly map the haloes with $\tau_\mathrm{pos}\leq 0.05$ and $\tau_\mathrm{mass}\leq 0.05$, and take them out from our halo catalogues. We then gradually increase the threshold values of $\tau_\mathrm{pos}$ and $\tau_\mathrm{mass}$, map the haloes in the remaining halo catalogues which satisfy the new conditions, and remove them from the halo catalogue. With the maximum threshold values of $\tau_\mathrm{pos,th} = 2.0$ and $\tau_\mathrm{mass,th}=0.6$, $\sim 90\%$ of haloes in the low resolution halo catalogue can be mapped into the high resolution ones. To test the effects of mis-mapping, we have varied the maximum threshold values and found that our conclusion in the following doesn't change.

The spin parameters $\lambda$ are calculated and used to obtain the ratio parameter
\begin{equation}\label{aeqn04}
r_\lambda = \frac{\lambda_n}{\lambda_{512}},
\end{equation}
where $n=64, 128, 256$ for $N_p = 64^3, 128^3, 256^3$ simulations respectively.

\begin{figure}
\includegraphics[width=500pt]{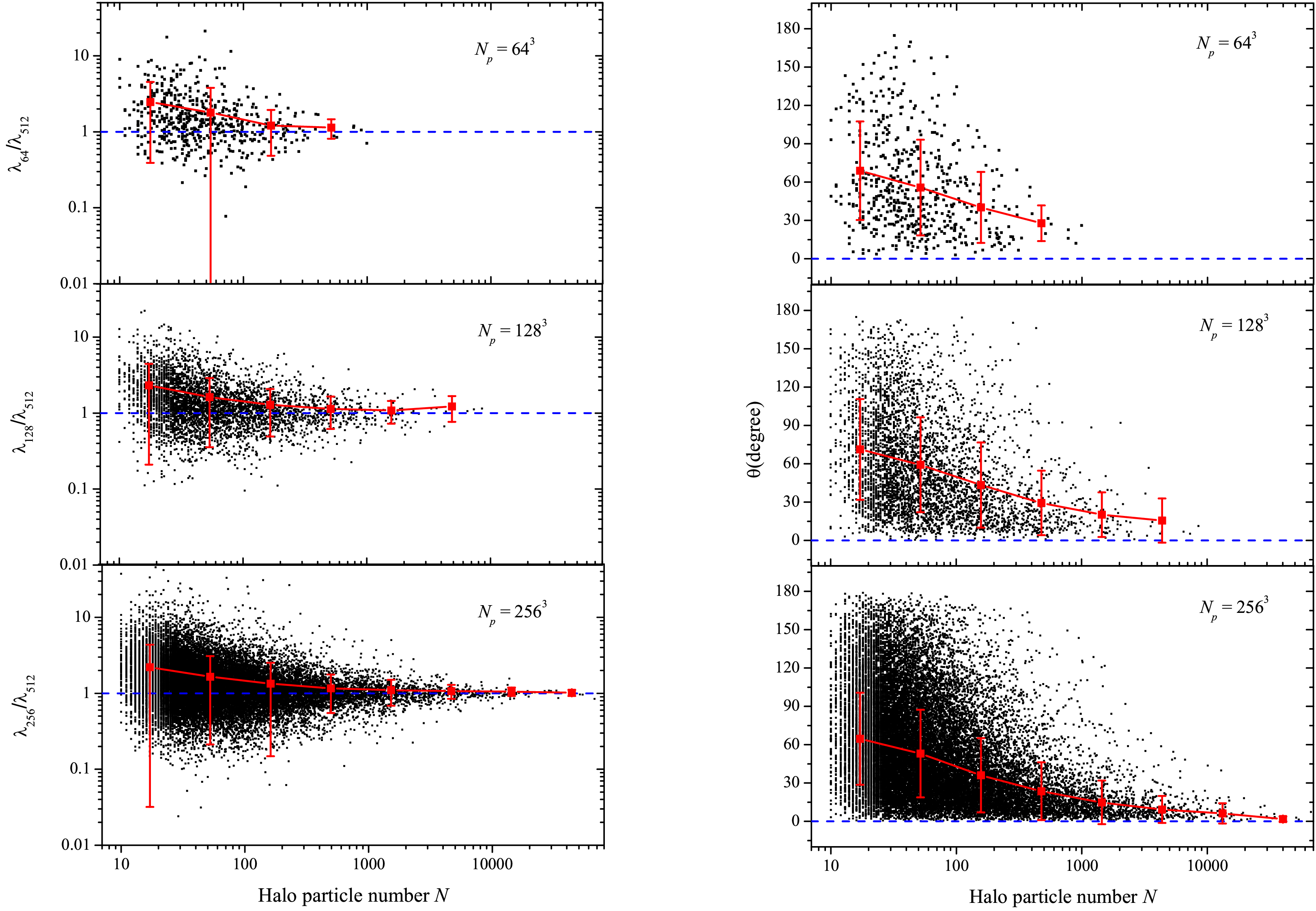}
\caption{Relation between $r_\lambda = \lambda_n/\lambda_{512}$ and halo particle number $N$ (left panel), and the relation between $\theta$ (the angle between ${\bf J}_n$ and ${\bf J}_{512}$) and halo particle number $N$ (right panel) for different low resolution simulations at $z=0$. The square dots are the mean of $r_\lambda$, or $\theta$ in different bins, and the associated error bars are $1\sigma$ standard deviations.} 
\label{f13}
\end{figure}

The dependence of $r_\lambda$ on halo particle number $N$ (or mass) is shown in Figure \ref{f13} (left panel). Low resolution haloes (small $N$) have a trend to overestimate the magnitudes of angular momenta. In particular, haloes with $\sim 20$ particles have an average overestimate of $r_\lambda \approx 2.0$. This is a numerical artifact and we can use it to find a suitable $N_\mathrm{min}$. To give an average estimate of the spin parameter within $20\%$ accuracy level, $N_\mathrm{min}\approx 200$ is required. In our $\Lambda$CDM512b simulations, we conservatively use haloes with $N_\mathrm{min}=400$ to study the $J-M$ relation (within $10\%$ accuracy level).

Although only the magnitudes of angular momenta are considered in the $J-M$ relation, as a reference and for interest, we also present the direction dependence on the halo resolution in Figure \ref{f13} (right panel), where $\theta$ is the angle between ${\bf J}_n$ and ${\bf J}_{512}$. For haloes with $N\geq 200$, on average, the directions of angular momenta in high and low resolutions agree with each other to within $45$ degrees. For haloes with $N\geq 400$, on average $\theta\leq 30$ degrees is obtained. 

\begin{figure}
\centering
\includegraphics[width=240pt]{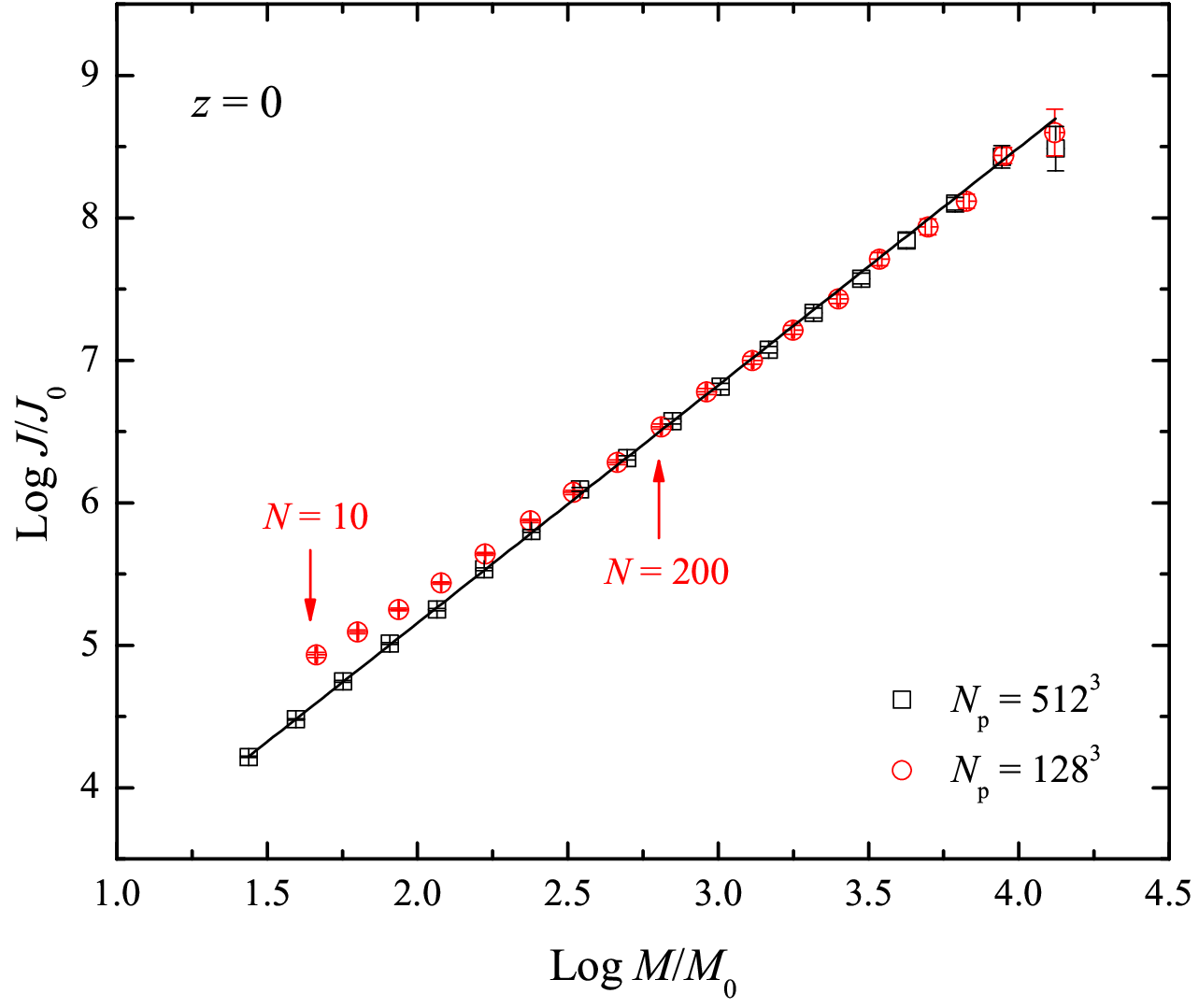}
\caption{Halo resolution effects on the $J-M$ relation. The data points and error bars here are similar to Figure \ref{f11}, which are the mean and standard deviations in the mass bins. The open squares (circles) show the $J-M$ relation from the $N_p = 512^3 (N_p=128^3)$ simulation. The solid line is the best-fit for the $J-M$ relation in the $N_p = 512^3$ simulation using MBF. Here $N_\mathrm{min} = 400$ for the $N_p = 512^3$ simulation. The arrows mark the corresponding halo particle number $N$ for $N_p=128^3$ simulations in two mass bins.} 
\label{f14}
\end{figure}

The resolution effects on the $J-M$ relation are shown in Figure \ref{f14}. Low resolution haloes tend to bend the $J-M$ relation upwards due to the fact that low resolution haloes overestimate the magnitudes of angular momenta. Haloes with $N\geq 200$ tend to give a converged $J-M$ relation as compared to high resolution simulations.

\section{Tests of fitting methods}\label{as2}

When fitting the $J-M$ relation, we use two independent methods: (1) All Points Fitting (APF). For every simulated realization, we performed a linear least square fitting for all data points in the $\log J-\log M$ plane, as most of scaling relation studies did. The final results were obtained from the mean and standard deviation among all realizations. (2) Mass Bins Fitting (MBF). In this method, we firstly divided data points into several equal-sized bins according to haloes' logarithmic masses. Then, for each bin, we calculated the mean values and standard deviations of both $\log J$ and $\log M$. For those bins with small number of data points, we use the bootstrap sampling method to get a better estimation. Finally, we used the total least square fitting method \citep{krystek2007} to fit the $J-M$ relation from the bin means by setting weights as reciprocal of squared bin standard deviations. Like the APF method, the final results were obtained from averaging over all realizations.

\begin{figure*}
\centering
\includegraphics[width=450pt]{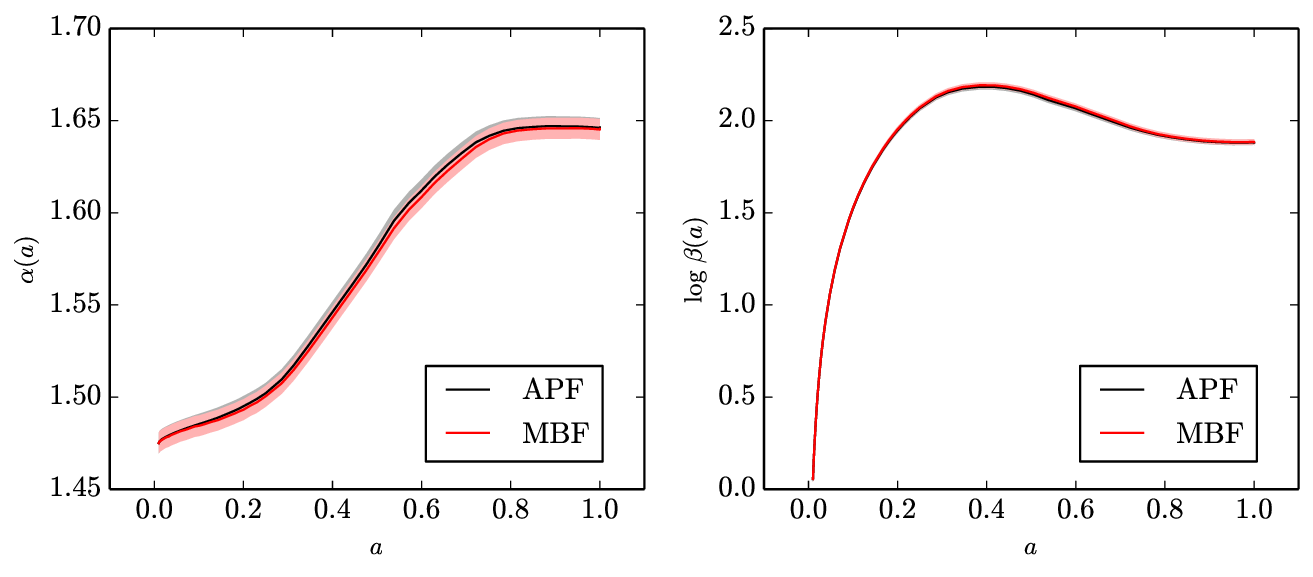}
\caption{Fitting method dependence of the $J-M$ relation. The left and right panel compare the fitted $\alpha$ and $\beta$ respectively for two fitting methods, APF (black) and MBF (red).}
\label{f12}
\end{figure*}

Examples of APF and MBF can be found in Figure \ref{f02} and Figure \ref{f11}. Comparison between these fitting methods is shown in Figure \ref{f12}. The maximum differences are smaller than $0.5\%$ for both $\alpha$ and $\log \beta$. We conclude that they give consistent fitting results. 

\section{Smoothing in TTT}\label{as4}
In this appendix, we test the smoothing method used in TTT. A smoothing process [Equation (\ref{eqn12})] is necessary in TTT for two reasons:

Firstly, as pointed out by \citet{white1984}, the validity of the Zel'dovich approximation requires $|\delta^2| < 1$. However, inside a protohalo region, there may exist some smaller scale perturbations with $|\delta^2| > 1$. To keep $|\delta^2| < 1$ for all scales within a protohalo during the whole period before the turnaround, we need to smooth perturbations with a scale equal to the protohalo scale. 

\begin{figure}
\centering
\includegraphics[width=240pt]{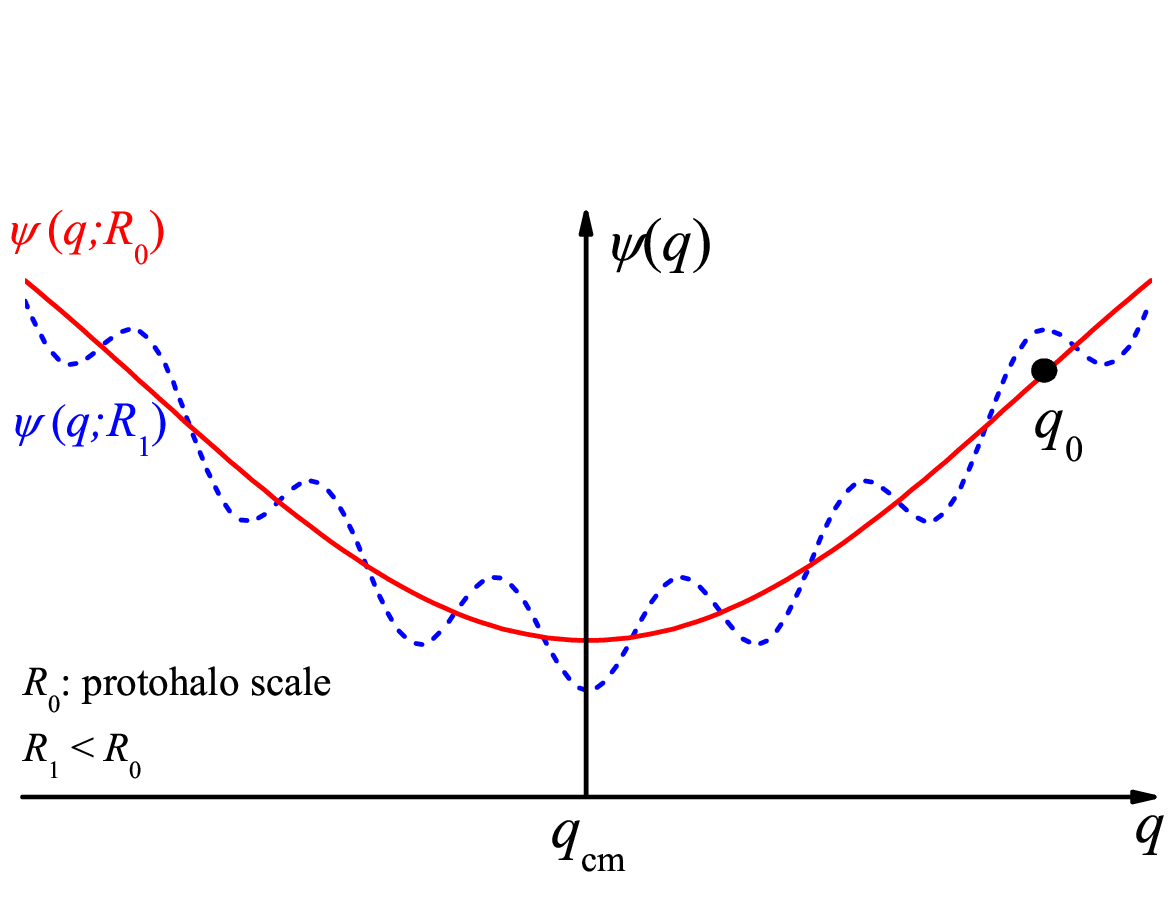}
\caption{Schematic plot of smoothed potentials with different smoothing scales $R_0$ (solid) and $R_1$ (dashed), with $R_1<R_0$.}
\label{f15}
\end{figure}

Secondly, truncating the Taylor expansion at second order requires that the smoothing scale $R$ of a protohalo should be comparable to its size. As shown in Figure \ref{f15}, using a smaller smoothing scale $R_1$, we can see more smaller hills and valleys which lead to failure of the Taylor approximation at certain points such as $q_0$. Similarly, one can expect that too large a smoothing scale is not acceptable either. Only with a smoothing scale comparable to the protohalo's size ($R=R_0$) can one approximate $\psi({\bf q})$ better within the whole protohalo.

In practice, a top-hat smoothing function with scale $R_0=(3M/4\pi\rho_0)^{1/3}$ is often used. Here, we test this smoothing scheme by comparing the results with different smoothing scales: $R=fR_0$, $R=R_0^m$ and a globally constant smoothing scale $R_\mathrm{global}$. Note that $R_0$ is the scale of a protohalo in the Lagrangian space, which depends only on the mass of the final virialized halo, and thus it is not a function of redshift. In the following, we use the particle distribution at $z=100$ (the initial condition) to test the prediction of TTT because the Zel'dovich approximation is expected to hold at high redshifts.

\begin{figure*}
\includegraphics[width=500pt]{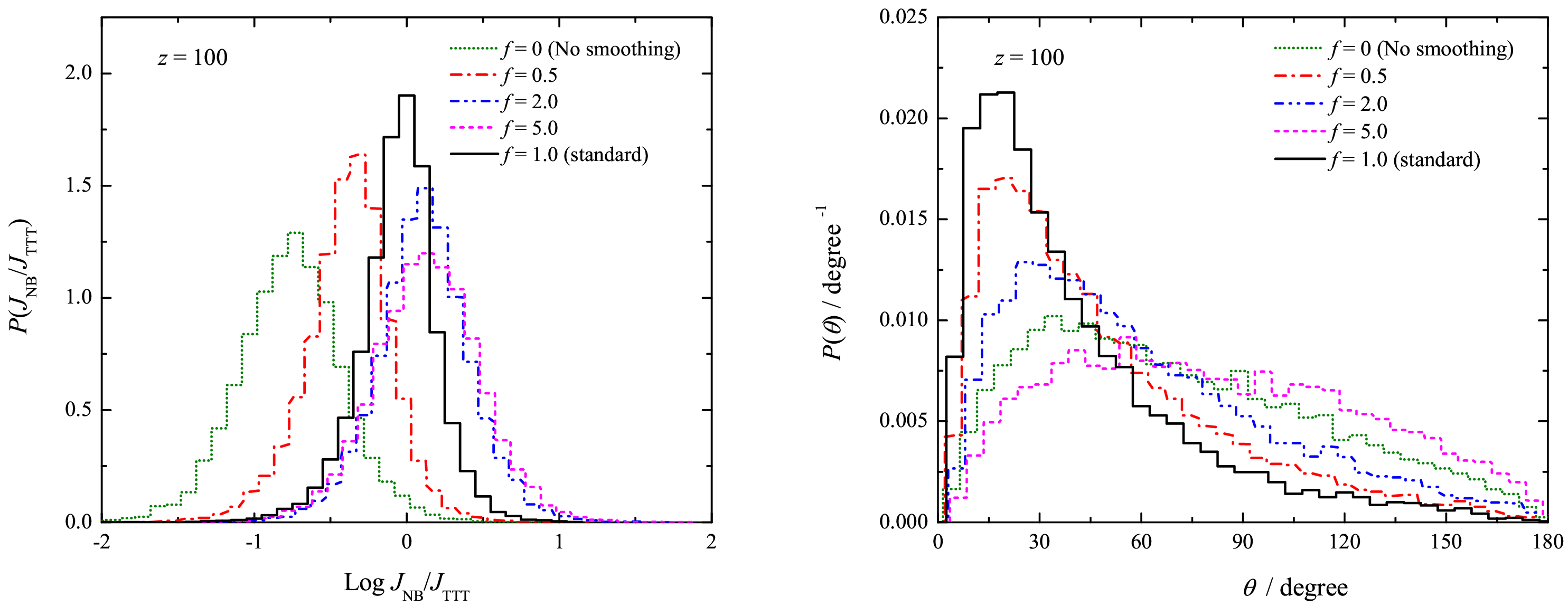}
\caption{TTT predictions with different smoothing scales $R=fR_0$ at the linear regime ($z=100$).}
\label{f16}
\end{figure*}

The probability distribution of $J_\mathrm{NB}/J_\mathrm{TTT}$ and $\theta$ for $R=fR_0$ are shown in Figure \ref{f16}, where $J_\mathrm{NB}$ is a protohalo's angular momentum measured from N-body simulation [Equation (\ref{eqn22})], $J_\mathrm{TTT}$ is the angular momentum predicted by TTT [Equation (\ref{eqn09})] and $\theta$ is the angle between ${\bf J}_\mathrm{NB}$ and ${\bf J}_\mathrm{TTT}$. With a smaller smoothing scale ($f<1.0$), TTT overestimates the magnitude of angular momentum and gives a poorer prediction of the spin direction. Using a larger scale ($f>1.0$), TTT underestimates the angular momentum magnitude and fails to predict its direction. More results can be found in Table \ref{t06}, together with the results for $R=R_0^m$.  

\begin{table*}
 \centering
  \caption{TTT predictions of the spin magnitude and direction with different smoothing scales $R=fR_0$ and $R=R_0^m$ (at $z=100$).}
  \begin{tabular} {@{}ccccccc@{}}
  \hline
     &   & $J_\mathrm{NB}/J_\mathrm{TTT}$ & & & $\theta$ (degree) & \\
 \hline
 $f$ & Mean & Median & Dispersion\footnote{The dispersions in this table are the 68.3\% confidence intervals. It's calculated by $\sigma_x = \frac{x_{84.2\%}-x_{15.9\%}}{2}$, where $x_{84.2\%}$ and $x_{15.9\%}$ are the values corresponding to $84.2\% $ and $15.9\%$ in the cumulative distribution of $x$.} & Mean & Median & Dispersion \\
 \hline
   0 & 0.23 & 0.17 & 0.13 & 69.8 & 63.3 & 45.1 \\
 0.2 & 0.32 & 0.24 & 0.18 & 61.0 & 52.2 & 41.7 \\
 0.5 & 0.48 & 0.40 & 0.24 & 47.3 & 37.5 & 33.9 \\
 0.8 & 0.72 & 0.63 & 0.32 & 39.3 & 29.4 & 28.6 \\
 0.9 & 0.81 & 0.71 & 0.36 & 39.0 & 28.9 & 28.6 \\
 1.0 & 0.93 & 0.82 & 0.42 & 38.7 & 28.7 & 28.6 \\
 1.1 & 1.04 & 0.91 & 0.47 & 38.8 & 28.7 & 28.6 \\
 1.2 & 1.16 & 1.01 & 0.54 & 39.3 & 28.9 & 29.2 \\
 1.5 & 1.38 & 1.14 & 0.67 & 44.7 & 35.0 & 32.2 \\
 2.0 & 1.45 & 1.11 & 0.79 & 57.7 & 48.8 & 39.3 \\
 5.0 & 1.59 & 1.14 & 0.96 & 78.2 & 74.1 & 47.6 \\              
\hline
 $m$ & Mean & Median & Dispersion & Mean & Median & Dispersion \\
 \hline
 0.5 & 0.40 & 0.33 & 0.22 & 52.9 & 43.4 & 38.8 \\ 
 1.2 & 1.37 & 1.13 & 0.67 & 44.7 & 34.9 & 32.5 \\
 \hline 
\end{tabular}
\label{t06}
\end{table*}

\begin{figure}
\centering
\includegraphics[width=240pt]{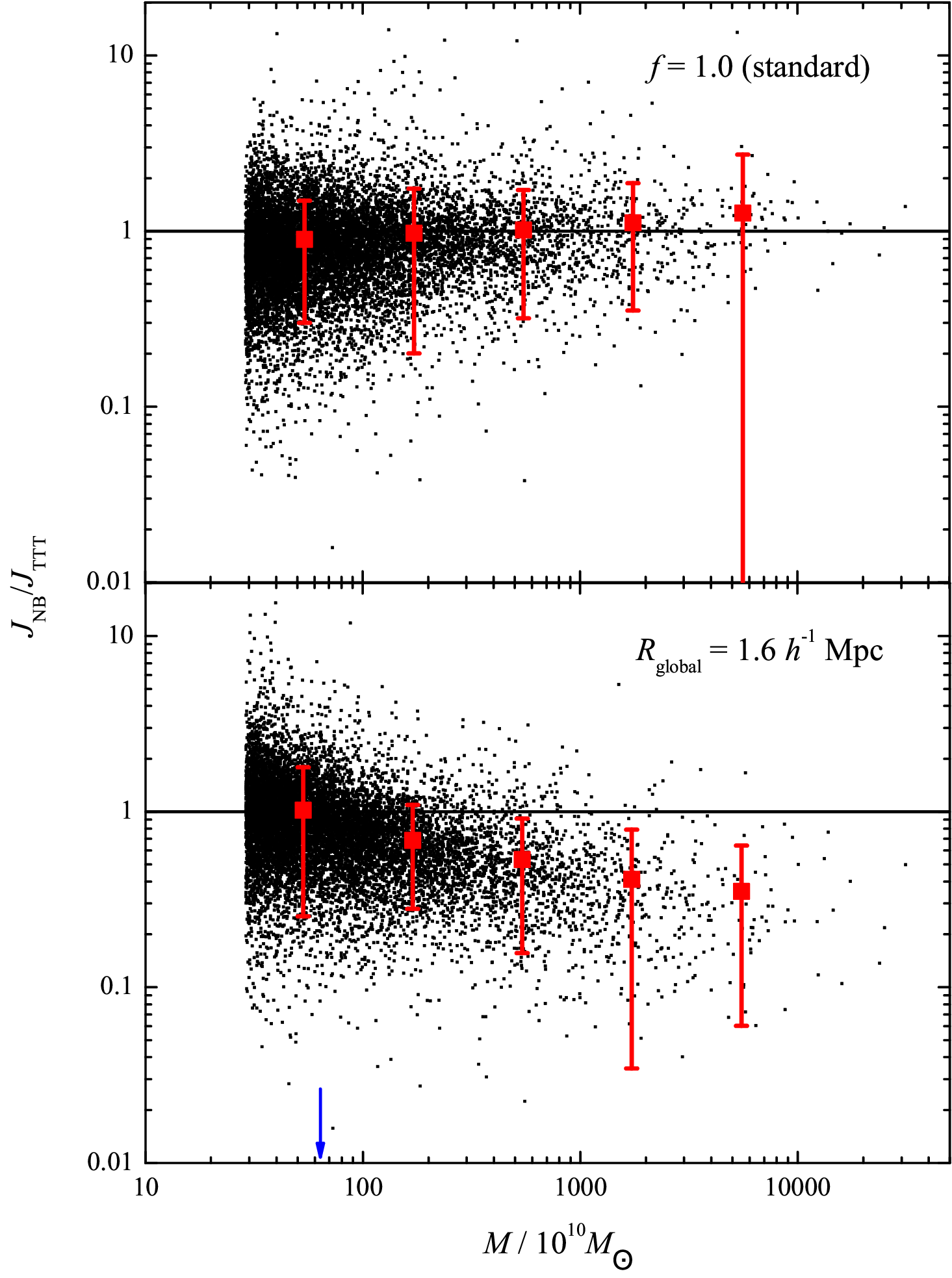}
\caption{Comparison of $J_\mathrm{NB}/J_\mathrm{TTT}$ for the standard choice of the smoothing scale ($R=R_0$) and a globally constant $R_\mathrm{global}=1.6 h^{-1}$ Mpc (at $z=100$). We plot $J_\mathrm{NB}/J_\mathrm{TTT}$ for all protohaloes (dark points) and the corresponding mean (solid square) and standard deviation (error bar) in each mass bin. The arrow marks the corresponding mass scale of $R_\mathrm{global}$.}
\label{f17}
\end{figure}

If we use a globally constant smoothing scale $R_\mathrm{global}$ (for example, $R_\mathrm{global}$ is the median length scale in our protohalo sample), then TTT overestimates (underestimates) the angular momenta of high (low) mass protohaloes, as shown in Figure \ref{f17}. Therefore, a globally constant smoothing scale is not suitable for TTT, either.

Among all these smoothing schemes, $R=R_0$ works best. We conclude that the smoothing of the potential is a key ingredient for TTT. Without potential smoothing, TTT fails to give acceptable predictions (see Figure \ref{f16} and Table \ref{t06} for the case of $f=0$). The smoothing scale $R=R_0$ introduces an additional mass dependence into TTT's predicted angular momentum and leads to a deviation of $\alpha$ from $5/3$ in the linear regime.

\end{document}